\documentclass[11pt,a4paper]{article}
\pdfoutput=1
\usepackage{ifpdf}
\usepackage{jheppub}
\usepackage{rotating}
\usepackage[bb=boondox]{mathalfa}
\usepackage{comment}

\usepackage{tikz-cd}
\usepackage[normalem]{ulem}
\usepackage{graphics} % Required for inserting images
\usepackage{mathrsfs}
\usepackage{amsmath,amsfonts,amssymb}
\usepackage{mathtools}
\usepackage{dcolumn}% Align table columns on the decimal point
\usepackage{bm}% bold math
\usepackage[mathlines]{lineno}% Enable numbering of text and display math
\usepackage{float}
\usepackage{blindtext}
\usepackage{titlesec}
\title{Sections and Chapters}
\usepackage[toc,page]{appendix}
\graphicspath{ {./images/} }
\usepackage[thinc]{esdiff}
\usepackage[euler]{textgreek}
\usepackage{braket}
\usepackage{physics}
\usepackage{xfrac}
\usepackage{soul}
\usepackage{stmaryrd}
\usepackage{pifont}
\usepackage{subcaption}
\usepackage{caption}
\usepackage{slashed}
\usepackage{ulem}
\usepackage{orcidlink}
\usepackage{booktabs}

\title{\textcolor{black}{Revisiting Singlet Fermion Dark Matter with a Scalar Portal: Connecting Higgs Phenomenology and Strong Electroweak Phase Transition }}
\author[a]{Jaydeb Das,\,\orcidlink{0000-0001-6335-9377}\,}
\author[b]{\!\!\,\,Saurabh Niyogi,\,\orcidlink{0009-0008-2355-8847}\,}
\author[c,d]{Tripurari Srivastava\,\orcidlink{0000-0001-6856-9517}\,}
\affiliation[a]{Department of Physics, Indian Institute of Technology Guwahati,
North Guwahati, Assam-781039, India.}
\affiliation[b]{Department of Physics, Gokhale Memorial Girls' College, Kolkata, West Bengal-700020, India.}

\affiliation[c]{Department of Physics and Astrophysics, University of Delhi, Delhi-110007, Delhi, India.} 
\affiliation[d]{Institute of Particle Physics and Key Laboratory of Quark and Lepton Physics (MOE),
Central China Normal University, Wuhan, Hubei 430079, China.}
\emailAdd{jaydebphys@rnd.iitg.ac.in}
\emailAdd{saurabhphys@gmail.com}
\emailAdd{tripurarisri022@gmail.com}

\definecolor{darkgreen}{cmyk}{1,0,1,0.4}
\definecolor{darkcyan}{cmyk}{1,0,0,0.4}
\allowdisplaybreaks

\abstract{We investigate a minimal extension of the Standard Model with a real singlet scalar and a singlet Dirac fermion acting as dark matter. Unlike a conventional singlet scalar setup, we assume that the singlet scalar does not acquire a vacuum expectation value at zero temperature. This decouples the scalar mixing angle from the Higgs–portal quartic coupling responsible for the strong first-order electroweak phase transition, allowing it to coexist with current collider and direct-detection constraints. The Higgs–singlet mixing is generated independently through a trilinear portal interaction. We check theoretical consistency conditions, various LHC limits on heavy scalar resonances, dark matter relic abundance, and direct detection bounds to delineate the viable parameter space. We perform a detailed analysis of the electroweak phase transition and show that a strong first-order transition is realized for a selected set of benchmark points. We further compute the resulting stochastic gravitational wave spectra and find that several scenarios yield signals potentially observable at future space-based interferometers. Our results establish a unified and testable framework that connects collider phenomenology, first-order electroweak phase transition, and the resulting production of gravitational waves, along with the dark matter phenomenology, all within a simple renormalizable extension of the Standard Model.}
\keywords{Dark matter, Electroweak phase transition, Gravitational wave, Collider phenomenology}

\begin{document}
\maketitle
\flushbottom

%\preprint{}

\section{Introduction}
The Standard Model (SM) of particle physics, despite its remarkable success in
describing the known fundamental particles and their interactions, remains
incomplete. In particular, it fails to account for the observed baryon
asymmetry of the Universe, namely the excess of matter over antimatter.
Electroweak baryogenesis offers an attractive explanation for this asymmetry,
provided that the electroweak phase transition (EWPT) is strongly first order.
However, for the experimentally measured Higgs boson mass of
$125~\mathrm{GeV}$, the electroweak phase transition in the SM proceeds as a
smooth crossover~\cite{Kajantie:1996mn,Huet:1994jb,Csikor:1998eu}, precluding the
strong out of equilibrium dynamics required to satisfy the Sakharov
conditions~\cite{Sakharov:1967dj}. Consequently, electroweak baryogenesis is
not viable within the SM framework~\cite{Anderson:1991zb,Morrissey:2012db},
motivating the exploration of physics beyond the SM (BSM). Among the many possible
extensions, minimal and economical scenarios that can address these
shortcomings without introducing unnecessary complexity are particularly
appealing.

Another major missing piece in our understanding of fundamental physics is the
nature of dark matter (DM), whose existence is firmly established by astrophysical
and cosmological observations but remains unexplained within the SM. A simple and widely studied possibility is to extend the SM by a real
gauge singlet scalar, which can itself serve as a dark matter candidate when
stabilized by a discrete symmetry~\cite{Cline:2013gha,GAMBIT:2017gge}. In such
scalar dark matter models, however, the stabilizing symmetry forbids
interactions involving odd powers of the singlet field. As a result, the
observed relic abundance is typically achieved only in the Higgs funnel
region, requiring a very small Higgs portal coupling. While this suppresses
direct detection rates, it also renders the scalar sector ineffective in
inducing a strong first-order electroweak phase transition
\cite{Profumo:2007wc,Noble:2007kk}. Increasing the portal coupling to strengthen
the phase transition is, on the other hand, tightly constrained by
direct detection experiments~\cite{XENON:2018clg,Arcadi:2019lka}. Consequently,
scalar dark matter extensions often face a tension between obtaining the
correct relic density and realizing a strong first-order electroweak phase
transition simultaneously.

A minimal way to alleviate this tension is to separate the roles of dark
matter and the electroweak phase transition. In this work, we consider an
extension of the SM containing a real singlet scalar and a singlet Dirac
fermion. The fermion constitutes the dark matter candidate, while the scalar
field modifies the Higgs potential and plays a central role in the dynamics of
the electroweak phase transition. The Yukawa interaction between the scalar
and the fermion governs the thermal freeze out of dark matter, whereas both
the scalar self interactions and the scalar fermion coupling contribute to
the finite temperature effective potential. This minimal field content
therefore provides a unified framework in which dark matter phenomenology and
electroweak phase transition dynamics are intrinsically linked, while leaving
distinct imprints in direct detection experiments, collider observables, and
cosmology.

In many singlet extended models, the scalar field acquires a vacuum
expectation value (VEV) at zero temperature, and achieving a strong first-order
EWPT requires a large Higgs portal coupling
\cite{Craig:2014lda,Robens:2015gla}. This simultaneously induces a large
Higgs singlet mixing angle, which is strongly constrained by LHC measurements
and direct detection bounds. In contrast, we adopt a different strategy by
requiring that the singlet scalar does not acquire a vacuum expectation value
at zero temperature~\cite{Profumo:2014opa,Beniwal:2017eik}. The Higgs singlet
mixing is instead controlled by an independent trilinear portal interaction.
This construction decouples the mixing angle from the quartic portal coupling,
allowing the latter to be large enough to generate a strong first-order
electroweak phase transition (SFOEWPT) while keeping the mixing angle small and
consistent with existing experimental constraints.

Though SFOEWPT can be realized in a variety of new-physics scenarios, including models with extended scalar sectors such as two Higgs doublet models (2HDMs)~\cite{Cline:1996mga,Bhatnagar:2025jhh}, various singlet or multi-scalar extensions~\cite{DiazSaez:2021pmg,Ghorbani:2021rgs,Ellis:2022lft,Ghorbani:2018yfr,Barger:2007im,Carena:2019une,Espinosa:2011ax,Noble:2007kk,Cline:2012hg,Alanne:2014bra,Cline:2013gha,Srivastava:2025oer,Chaudhuri:2022sis,Borah:2024emz,Chiang:2019oms,Kang:2017mkl,Kannike:2019mzk,Ghorbani:2024twk,Ghorbani:2019itr,DiazSaez:2024nrq,DiazSaez:2021pfw,Murai:2025hse,Chakrabarty:2022yzp,Ferber:2023iso}, or higher-dimensional operators~\cite{Anderson:1991zb,Zhang:1992fs,Camargo-Molina:2021zgz,Hashino:2022ghd,Oikonomou:2024jms,Gazi:2024boc}; we stick to a minimal extension of the SM. Thus, such minimal particle content beyond the SM within the renormalizable and gauge-invariant framework is not only able to explain the observed dark matter relic abundance, but also modifies the structure of the scalar potential in such a way that opens the possibility of a strong first-order phase transition, which is one of the required conditions for baryogenesis in the present universe. In addition, the generation of detectable stochastic gravitational waves from the expansion and subsequent collision of bubbles of the broken phase offers another experimental channel to test such a scenario \cite{Kamionkowski:1993fg,Ellis:2018mja,Croon:2018erz,Beniwal:2018hyi,Huang:2017laj,Hashino:2018zsi,Demidov:2017lzf,Mazumdar:2018dfl,Kobakhidze:2016mch,Kobakhidze:2017mru,Dev:2016feu}. The breakthrough discovery by LIGO~\cite{Abbott:2016blz} has opened a new window into early universe dynamics, motivating the study of electroweak-scale GWs as a complementary probe of new physics in collider. The amplitude and peak frequency of the resulting GW spectrum encode information about the underlying scalar potential, making future space-based interferometers—such as LISA~\cite{eLISA:2013xep}, DECIGO~\cite{Kawamura:2011zz}, BBO~\cite{Corbin:2005ny}, TAIJI~\cite{Gong:2014mca}, and TianQin~\cite{TianQin:2015yph}—promising tools for testing SFOEWPT scenarios. Therefore, the mixing angle between the SM doublet and real singlet that appears in the Higgs couplings controls the phenomenology of the heavier scalar state. On top of that, the new scalar interactions allow for non-resonant and resonant collider signatures - thus offering a well-defined and testable link between cosmological considerations and collider experiments.

The structure of the paper is as follows. In Section~\ref{sec:model}, we
introduce the theoretical setup of the model. Section~\ref{sec:constraints}
is devoted to theoretical, collider, and electroweak precision constraints.
The dark matter phenomenology, including relic density and direct detection,
is discussed in Section~\ref{sec:darkmatter}. In Section~\ref{sec:ewpt}, we
analyze the electroweak phase transition and the associated gravitational wave
signals. Finally, we summarize our results and conclude in
Section~\ref{sec:conclusion}.

%%%%%%%%
%%%%%%%%%%% THIS PART IS REWRITTEN BY JAYDEB ON 18 NOVEMBER

\section{The framework}\label{sec:model}
We extend the SM by introducing a real gauge-singlet scalar field $s$ and a Dirac fermion $\chi$ which is also a singlet under the SM gauge group. In this framework, the fermion $\chi$ plays the role of a  WIMP-like (weakly interacting massive particle) DM. To guarantee its stability, we impose a discrete $\mathbb{Z}_2$ symmetry under which $\chi$ is odd, while the scalar field $s$ and all SM fields remain even. This symmetry forbids any interaction of $\chi$ directly with the SM particles. Consequently, $\chi$ can not decay directly into any SM pair. 
Therefore, the most general renormalizable Lagrangian consistent with the SM gauge symmetries and the imposed $\mathbb{Z}_2$ parity for this scenario can be written as:
\begin{equation}\label{eq:Lag}
\mathcal{L} \supset \frac{1}{2}(\partial_\mu s)^2 + \bar{\chi}(i\!\not{\partial}-m_\chi)\chi + \mathcal{L}_{\chi s} - V(H,s),
\end{equation}
where $m_\chi$ is the bare mass term of $\chi$ and $\mathcal{L}_{\chi s}$ is the Lagrangian containing the scalar-DM Yukawa interaction. The most general renormalizable scalar potential $V(H,s)$ is~\cite{Profumo:2007wc,OConnell:2006rsp}
\begin{align}
V(H,s) =&\; \mu_h^2 |H|^2 + \lambda_h |H|^4 + \frac12 \mu_s^2 s^2 + \frac14 \lambda_s s^4 
+ \frac12 \lambda_{hs}|H|^2 s^2  \nonumber\\
& +\, \mu_{hs}|H|^2 s + \frac13 \mu_3 s^3 + a_1 s,
\end{align}
where the Higgs doublet at zero temperature is
\begin{equation}
H = \frac{1}{\sqrt{2}}
\begin{pmatrix}
\chi_1 - i\chi_2 \\
h + v + i\chi_3
\end{pmatrix}, 
\end{equation}
where $v \approx 246~\mathrm{GeV}$ denotes the VEV of the Higgs field at zero temperature. Since the DM candidate $\chi$ is stabilized by a $\mathbb{Z}_{2}$ symmetry under which $\chi \to -\chi$, the only renormalizable Yukawa interaction between the singlet scalar field and the fermion field is~\cite{Fairbairn:2013uta,Kim:2008pp}
\begin{equation}\label{eq:Yuk}
\mathcal{L}_{\chi s} = g_{\chi}\, s\, \bar{\chi}\chi ,
\end{equation}
where $g_{\chi}$ is a dimensionless coupling constant. 
%The imposed $\mathbb{Z}_{2}$ parity forbids any direct coupling of $\chi$ to the Standard Model fields, thereby ensuring its stability on cosmological timescales. 
Throughout this work, we assume that the singlet scalar does not have VEV at zero temperature~\cite{Kozaczuk:2019pet,Chen:2017qcz,Chen:2014ask}. Under this assumption, the bare mass parameter $m_\chi$ of the singlet fermion $\chi$ (see Eq.~\eqref{eq:Lag}) coincides with its physical mass.

\subsection*{Scalar potential at tree level}
Expanding the scalar potential $V(H,s)$ in terms of the background fields $(h,s)$ gives the tree-level potential
\begin{align}\label{eq:treepot}
V_0(h,s) =
\frac12 \mu_h^2 h^2 + \frac12 \mu_s^2 s^2
+ \frac14 \lambda_h h^4 + \frac14 \lambda_s s^4
+ \frac14 \lambda_{hs} h^2 s^2  
+ \frac12 \mu_{hs} h^2 s + \frac13 \mu_3 s^3 + a_1 s .
\end{align}
Since the BSM scalar is a real gauge-singlet field, a constant shift can be used to choose a basis in which its VEV vanishes, $v_s = 0$. On this basis, the singlet tadpole condition ensures that $(h,s) = (v,0)$ is a stationary point of the tree-level potential and fixes $a_1$ in terms of the remaining parameters~\cite{Kozaczuk:2019pet}. 
At $T=0$, the corresponding stationary conditions yield the following relations:
\begin{align}
\left. \frac{\partial V_0}{\partial h} \right|_{(v,0)} = 0
&\;\Rightarrow\;
\mu_h^2 = -\lambda_h v^2 , \\
\left. \frac{\partial V_0}{\partial s} \right|_{(v,0)} = 0
&\;\Rightarrow\;
a_1 = -\dfrac{\mu_{hs}\, v^2}{2} .
\end{align}

\subsection*{Scalar mixing matrix}
Although the singlet scalar field $s$ does not acquire a VEV at zero temperature, the term $\mu_{hs} s |H|^2$ induces mixing between the Higgs and the singlet scalar after the spontaneous breaking of the electroweak symmetry. The CP-even scalar mass matrix at $(v,0)$ is
\begin{equation}
M^2(v,0) =
\begin{pmatrix}
2\lambda_h v^2 & \mu_{hs}\, v \\
\mu_{hs}\, v & \mu_s^2 + \tfrac12 \lambda_{hs}\, v^2
\end{pmatrix}\equiv
\begin{pmatrix}
M_{hh}^2 & M_{hs}^2 \\
M_{hs}^2 & M_{ss}^2
\end{pmatrix}.
\end{equation}
To obtain the physical scalar masses, the mass matrix is diagonalized by an orthogonal transformation, yielding the mass eigenstates $h_{1}$ and $h_{2}$ defined as
\begin{equation}
\begin{pmatrix} h_1 \\ h_2 \end{pmatrix}
=
\begin{pmatrix}
\cos\theta & -\sin\theta \\
\sin\theta & \cos\theta
\end{pmatrix}
\begin{pmatrix} h \\ s \end{pmatrix},
\end{equation}
Here $h_{1}$ is identified with the observed $125~\mathrm{GeV}$ Higgs boson, and $\theta$ denotes the Higgs--singlet mixing angle. The gauge eigenstates can therefore be expressed in terms of the mass eigenstates as
\begin{eqnarray}
    h &=& \cos\theta\, h_{1} + \sin\theta\, h_{2}, \quad
    s = \cos\theta\, h_{2} - \sin\theta\, h_{1},
\end{eqnarray}
and the corresponding mass eigenvalues are
\begin{align}
m_{h_1}^2 &= M_{hh}^2 \cos^2\theta + M_{ss}^2 \sin^2\theta 
            - 2 M_{hs}^2 \sin\theta\cos\theta , \\
m_{h_2}^2 &= M_{hh}^2 \sin^2\theta + M_{ss}^2 \cos^2\theta 
            + 2 M_{hs}^2 \sin\theta\cos\theta .
\end{align}
Conversely, the parameters of the scalar potential may be expressed in terms of the physical inputs:
\begin{eqnarray}
&&\lambda_h = 
\dfrac{m_{h_1}^2 \cos^2\theta + m_{h_2}^2 \sin^2\theta}{2v^2}, \quad
\mu_{hs}  =
\dfrac{(m_{h_2}^2 - m_{h_1}^2)}{v}\sin\theta\cos\theta, \label{eq:muhs} \\
&&\mu_s^2   = 
m_{h_1}^2 \sin^2\theta + m_{h_2}^2 \cos^2\theta 
- \dfrac{\lambda_{hs}\, v^2}{2} .
\end{eqnarray}
These relations allow one to trade the potential parameters for the physical masses and the mixing angle. In summary, the independent input parameters of the scalars--DM sector for the model are:
\[
\{\, v,\; m_{h_1},\; m_{h_2},\; \sin \theta,\; g_{\chi},\; m_{\chi},\; \lambda_{hs},\; \mu_{3},\; \lambda_s \,\}.
\]
Here, we assume that the real singlet scalar does not acquire a VEV at zero temperature. Consequently, the Higgs--singlet mixing arises solely from the dimensionful interaction term $\mu_{hs}$ rather than from the Higgs-portal coupling $\lambda_{hs}$. As a result, the direct-detection cross section is independent of $\lambda_{hs}$. This allows us to choose a relatively large portal coupling to achieve a SFOEWPT, while remaining fully consistent with current direct-detection constraints, as discussed later.
%%%%%%%%%%%%%%%
%%%%%%%%%%%%%%%%

\begin{comment}
\end{comment}

%%%%%%%%%%%%%%%%% 

\section{Constraints}\label{sec:constraints}

In this section, we summarize the theoretical and experimental constraints relevant to the singlet--fermion plus singlet real--scalar extension of the SM. We begin with the requirements of theoretical consistency, followed by a discussion on constraints coming from electroweak precision observables (EWPOs). Finally, we present collider bounds on the heavier scalar state~$h_2$ along with the Higgs signal strength measurements.

\subsection{Theoretical constraints}

A consistent scalar potential must be bounded from below, perturbative, and admit the electroweak vacuum as a stable minimum.

\paragraph{(i) Vacuum stability.}
The quartic couplings must satisfy the positivity conditions
\begin{equation}
\lambda_h \ge 0, 
\qquad
\lambda_s \ge 0,
\qquad
\lambda_{hs} \ge -2\sqrt{\lambda_h\lambda_s},
\label{eq:vacstab}
\end{equation}
ensuring that the potential is bounded from below. For $\lambda_{hs} \ge 0$ these conditions are automatically satisfied.

\paragraph{(ii) Perturbative unitarity.}
Tree-level perturbative unitarity of scalar $2\to 2$ scattering amplitudes requires~\cite{Chiang:2020yym,Zhou:2020ojf,Kang:2013zba}
\begin{eqnarray}
&&|\lambda_h| \le 4\pi,
\qquad
|\lambda_s| \le 4\pi,
\qquad
|\lambda_{hs}| \le 8\pi ,\nonumber\\
&&\Big|\, 3\lambda_h + 2\lambda_s \;\pm\;
\sqrt{
(3\lambda_h
- 2\lambda_s)^2
+ 2\lambda_{hs}^2
}
\,\Big| \le 8\pi.
\label{eq:unitarity}
\end{eqnarray}

\subsection{Electroweak precision observables}

Mixing between the singlet scalar and the SM Higgs modifies electroweak precision observables, particularly the oblique parameters \(S\) and \(T\) and the \(W\)-boson mass. These effects grow with both the mixing angle \(\theta\) and the heavy-scalar mass \(m_{h_2}\) \cite{Lopez-Val:2014jva,Dawson:2021ofa}. In the heavy-scalar regime, \(m_{h_2}\gtrsim 800~\mathrm{GeV}\), electroweak precision data typically imply an upper bound on the mixing angle at the level of \( |\sin\theta| \lesssim 0.2 \). For the lighter mass range relevant to our analysis, \(m_{h_2}<600~\mathrm{GeV}\), the corresponding bound is weaker and usually remains at the level of a few times \(10^{-1}\), depending on the scalar mass \cite{Dawson:2021ofa}. This constraint is complementary to Higgs signal-strength limits and becomes more important for larger \(m_{h_2}\).

\subsection{LHC constraints on the heavy scalar \texorpdfstring{$h_2$}{h2}}
In the real scalar singlet extension considered here, electroweak symmetry breaking leads to
two physical CP-even scalars: the observed Higgs boson $h_1$ with mass $m_{h_1} = 125$ GeV
and a heavier scalar $h_2$ with mass $m_{h_2}$. The mixing in the scalar sector suppresses
all couplings of $h_1$ to SM quarks and gauge bosons by factor of $\cos \theta$. Similarly, the couplings to heavier scalar to SM quarks and gauge bosons also get scaled by $\sin \theta$.
\paragraph{Production rate.}
The production cross section of $h_2$ at the LHC is therefore given by
\begin{equation}
\sigma(pp\to h_2) \simeq \sin^2\theta \;\sigma_{\rm SM}(m_{h_2}),
\end{equation}
where $\sigma_{\rm SM}(m_{h_2})$ denotes the Standard Model Higgs production cross section evaluated at mass $m_{h_2}$.

\paragraph{Decay pattern.}
The branching ratios of $h_2$ follow those of an SM Higgs boson with mass $m_{h_2}$, up to kinematic differences such as the opening of the decay $h_2\to h_1 h_1$ for
$m_{h_2} > 2m_{h_1}$. We compute the signal rate
\begin{equation}
\sigma(pp\to h_2)\times {\rm BR}(h_2\to XX)
\end{equation}
across the $(m_{h_2},\sin\theta)$ parameter space using \texttt{MadGraph5\_aMC@NLO} \cite{Alwall:2011uj}. These predictions from the model are then compared to the 95\% C.L. upper limits reported by ATLAS and CMS in Run 2 searches for heavy scalar. The experimental limits are taken from the official HEPData tables and interpolated across the mass range shown in the figures. The three most relevant final states for heavy-scalar searches are:
\begin{itemize}
\item $h_2\to h_1 h_1$,
\item $h_2\to ZZ\to 4\ell$,
\item $h_2\to WW\to \ell\nu\ell\nu$.
\end{itemize}

\begin{figure}
\centering
\includegraphics[width=0.6\textwidth,height=6.5cm]{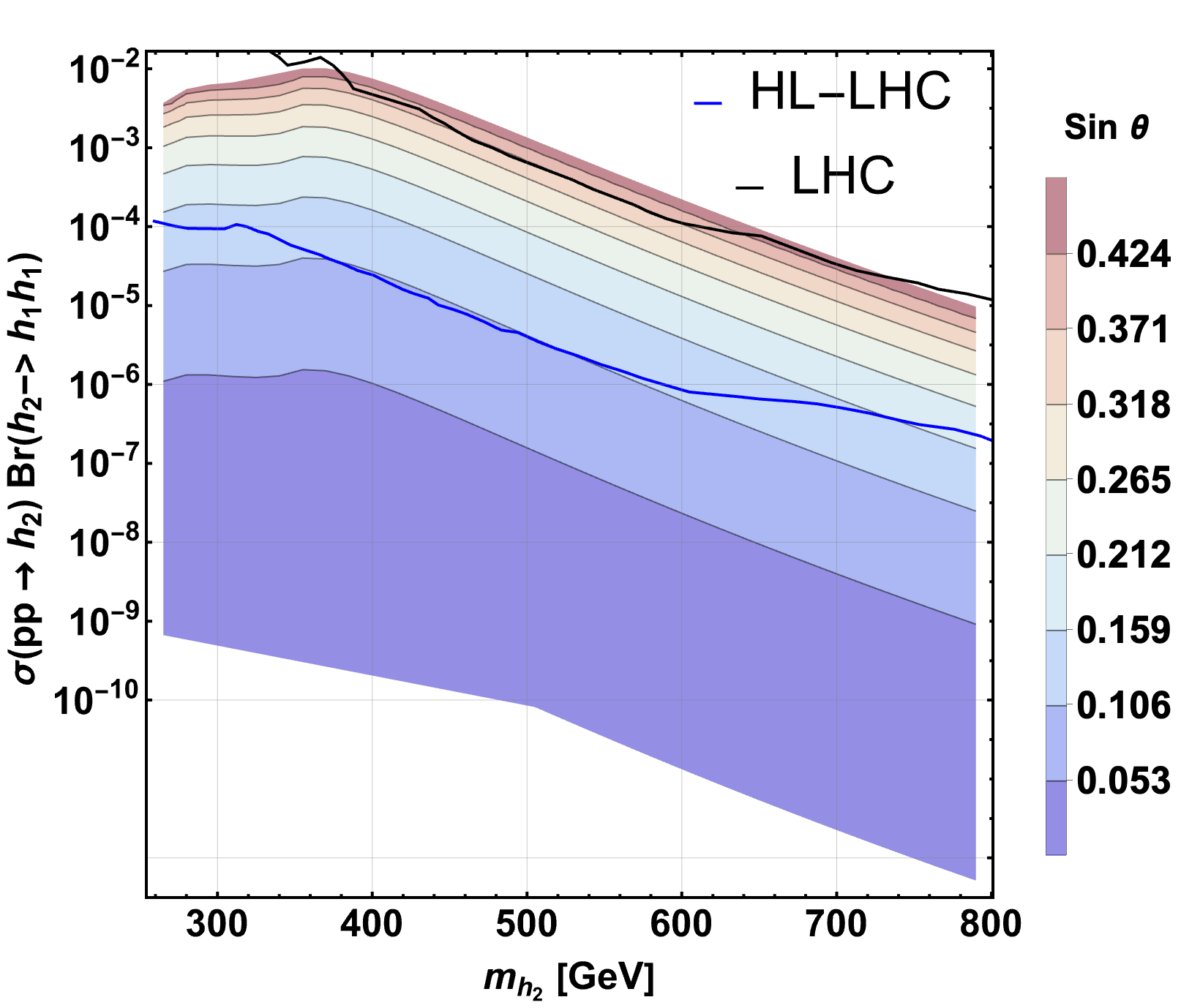}
\caption{
Predicted cross section $\sigma(pp\to h_{2})\times \text{BR}(h_{2}\to h_{1}h_{1})$
in the $(m_{h_{2}},\sin\theta)$ plane.  
The solid curve shows the ATLAS 95\% CL upper limit; the region above the curve is
excluded.}
\label{fig:h2hh}
\end{figure}

\paragraph{Di-Higgs final state.}
For scalar masses around \(m_{h_{2}} \simeq 460\text{--}600~\mathrm{GeV}\),
the decay \(h_{2} \rightarrow h_{1} h_{1}\) becomes increasingly important.
In this mass window, the ATLAS resonant di-Higgs searches~\cite{ATLAS:2021tyg}
provide their strongest constraint, allowing mixing angles only below
\[
\sin\theta \lesssim 0.36 .
\]
For \(m_{h_{2}} \gtrsim 600~\mathrm{GeV}\), the exclusion gradually weakens as
the production cross section and the \(h_{2}\to h_{1}h_{1}\) branching ratio
decrease.

At the High-Luminosity LHC, the di-Higgs sensitivity is expected to improve
substantially~\cite{ATL-PHYS-PUB-2025-018}. The projected limits strengthen the constraint on the mixing angle to roughly
\[
\sin\theta \lesssim 0.20\text{--}0.25,
\]
depending mildly on \(m_{h_{2}}\), thereby probing a significantly wider region
of parameter space.

Figure~\ref{fig:h2hh} displays the predicted di-Higgs yield, with the black
curve showing the present 95\%~CL exclusion from LHC searches and the blue
curve indicating the expected reach of the HL-LHC.

\paragraph{ZZ\,\texorpdfstring{$\to$}{->} 4$\ell$ channel.}
Due to the excellent invariant-mass resolution and the very small background 
in the \(ZZ \to 4\ell\) channel, the strongest constraint on the mixing angle 
is obtained in this mode.  Using the observed limits reported in 
Ref.~\cite{ATLAS:2020tlo}, the upper limit on the allowed mixing angle is
\begin{equation}
\sin\theta \lesssim 0.13- 0.20
\qquad
\text{for}\quad
250~{\rm GeV} \lesssim m_{h_2} \lesssim 700~{\rm GeV}.
\end{equation}
The bound becomes less strong as a result of the reduced production rate at higher masses.
\begin{figure}%
%    \centering
    \subfloat[]{{\includegraphics[width=0.5\textwidth,height=6.0cm]{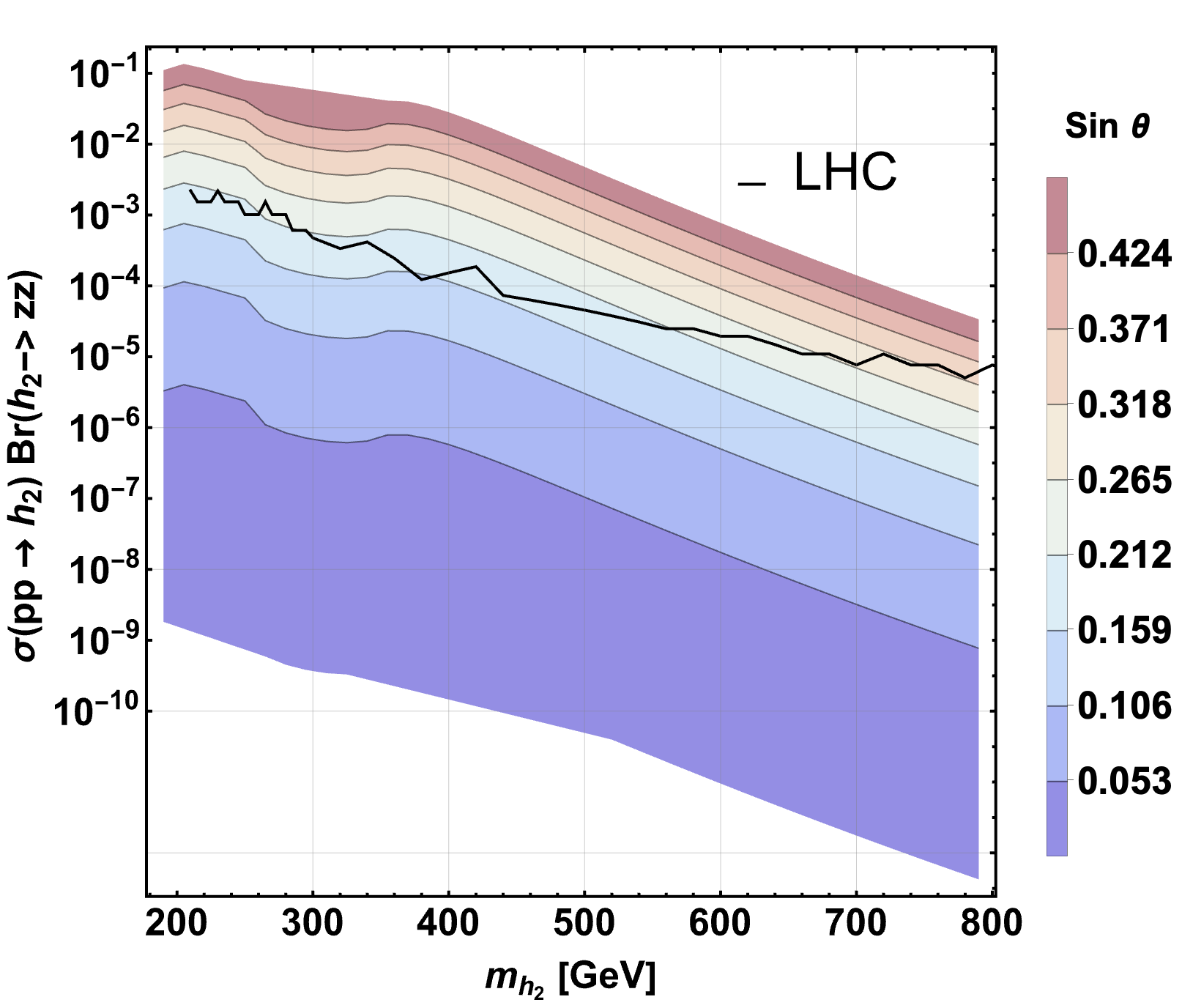} }}%
    \qquad
    \subfloat[]{{\includegraphics[width=0.5\textwidth,height=6.0cm]{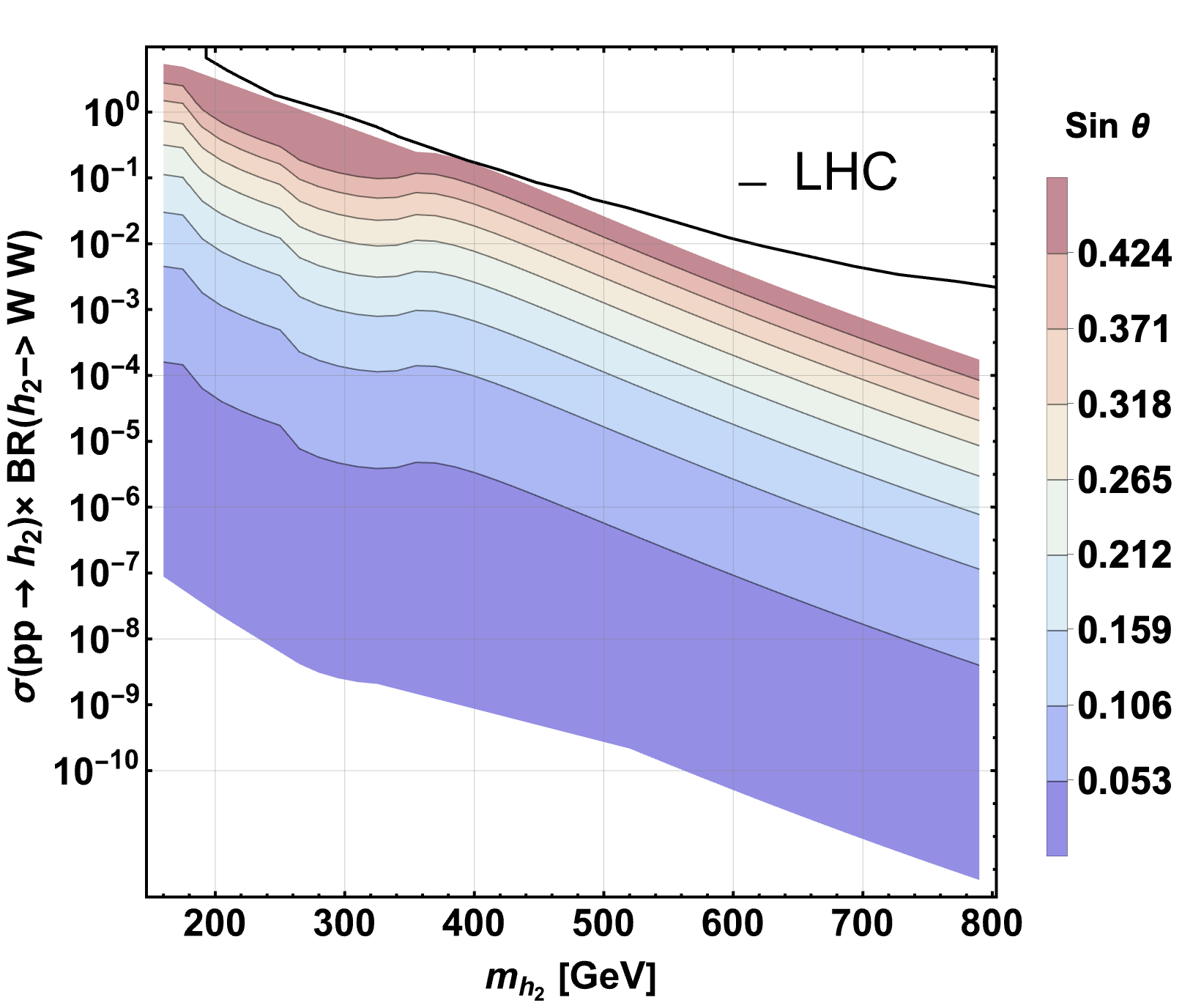} }}%
    \caption{ Exclusion in the $ZZ$ final state.  
The $ZZ\to 4\ell$ channel provides the strongest constraint across most of the mass
range. Constraints from the $WW$ final state.  The sensitivity is weaker than $ZZ$ due to larger backgrounds, but remains important for intermediate masses.}
\label{fig:h2WW}%
\end{figure}

\paragraph{WW channel.}
Although the branching ratio of the 125~GeV Higgs boson to \(WW\) is larger 
than that to \(ZZ\), the corresponding experimental analysis is affected by 
substantially higher backgrounds, resulting in a much weaker exclusion power.  
As reported in Ref.~\cite{ATLAS:2022enb}, the \(WW\) final state therefore 
provides only a modest constraint on the Higgs--singlet mixing angle, with 
values as large as \(\sin\theta \simeq 0.5\) remaining allowed across the full 
mass range considered. Consequently, this channel is far less sensitive than 
the \(ZZ\to 4\ell\) mode or the resonant di--Higgs searches as can be observed from left panel ($ZZ$ channel) and right panel ($WW$ channel) of Fig.\,\ref{fig:h2WW}. The exclusion further degrades at higher masses due to the usual decrease in the production cross section.

\paragraph{Combined collider limit.}
Collectively, heavy-scalar searches provide upper limit as
\begin{equation}
\sin\theta \lesssim \mathcal{O}(0.1)
\qquad
\text{for}
\quad
m_{h_2}\lesssim 600~{\rm GeV}.
\end{equation}

\subsection{Higgs signal strength constraint}

%Since $h_1$ couples as $\cos\theta$, the global signal strength scales as
Since \(h_1\) couples to SM states with a universal rescaling factor \(\cos\theta\), the corresponding Higgs signal strength modifier,
\[
\mu_{\rm sig}\equiv \frac{\sigma(pp\to h_1)\times {\rm BR}(h_1\to X)}
{\left[\sigma(pp\to h)\times {\rm BR}(h\to X)\right]_{\rm SM}},
\]
scales as
\begin{equation}
\mu_{\rm sig} = \cos^2\theta.
\end{equation}
ATLAS Run-II measurements yield~\cite{ATLAS:2022vkf} %\cite{Lane:2024yfr}
\begin{equation}
\mu_{\rm ATLAS} = 1.04 \pm 0.06,
\end{equation}
which implies, at the 95\% C.L.,
\begin{equation}
\sin\theta \lesssim 0.24.
\end{equation}
This constraint is independent of $m_{h_2}$ and can be stronger than direct heavy-scalar searches.

\subsection{Future HL-LHC sensitivity}

The High-Luminosity LHC (HL-LHC) is expected to achieve substantially improved precision in Higgs coupling measurements, leading to a projected bound~\cite{ATLAS:2025eii,Cepeda:2019klc,Banerjee:2020tqc}
\begin{equation}
|\sin\theta| \lesssim 0.17.
\end{equation}
Direct searches for a heavy scalar will probe mixings down to
\begin{equation}
|\sin\theta|\sim 0.05\text{--}0.10
\qquad
\text{for}
\quad
200~{\rm GeV}\lesssim m_{h_2}\lesssim 600~{\rm GeV}.
\end{equation}
These improvements will cover a large portion of the parameter space still allowed by present data.

\section{Dark matter study}\label{sec:darkmatter}

In this section, we study the phenomenology of a WIMP-like singlet fermionic dark matter. We compute the dark-matter relic abundance and the spin-independent DM–nucleon scattering cross section using \texttt{micrOMEGAs}~\cite{Alguero:2023zol,Belanger:2004yn}. In the following subsections, we present a detailed analysis of the relic-density constraints and direct-detection bounds, and identify the corresponding allowed regions of the parameter space.

\subsection{Relic density}

In this model, the singlet Dirac fermion $\chi$ serves as the DM candidate as mentioned earlier. It couples only with the real singlet scalar field $s$ via the Yukawa interaction given in Eq.~\eqref{eq:Yuk}. After electroweak symmetry breaking (EWSB) through the VEV of the Higgs field, its annihilation proceeds through the two scalar mediators $h_{1}$ and $h_{2}$. In addition to DM annihilation to SM particles, some other important channels like $h_i h_j~ (i,j=1,2)$ become crucial.  Hence, the abundance after thermal freeze-out is determined primarily by two
parameters (with the dominant one depending on the DM mass range): the scalar--Higgs mixing angle $\sin\theta$, which controls the mediator–SM interactions, and the Yukawa coupling $g_\chi$, which sets the strength of the vertices $\bar{\chi}\chi h_{1,2}$. Additionally, Higgs portal coupling $\lambda_{hs}$ and trilinear coupling $\mu_3$ play a crucial role in achieving the correct relic density. Throughout the entire parameter space, the relic density is independent of the scalar quartic coupling $\lambda_s$.

Depending on the relative sizes of $\sin\theta$, $g_\chi$ and $\lambda_{hs}$, the correct thermal relic 
density can be achieved in different (and sometimes mutually disjoint) regions of the 
parameter space. The viable regions can be grouped into three characteristic classes:
(i) the Higgs funnel $m_\chi \simeq m_{h_1}/2$, 
(ii) the singlet-scalar resonance $m_\chi \simeq m_{h_2}/2$, and 
(iii) the degenerate region where annihilation is enhanced by near-kinematic alignment lies beyond the $m_\chi \approx m_{h_2}$ region. 

\begin{figure}[ht!]
    \centering
    \includegraphics[width=0.49\linewidth]{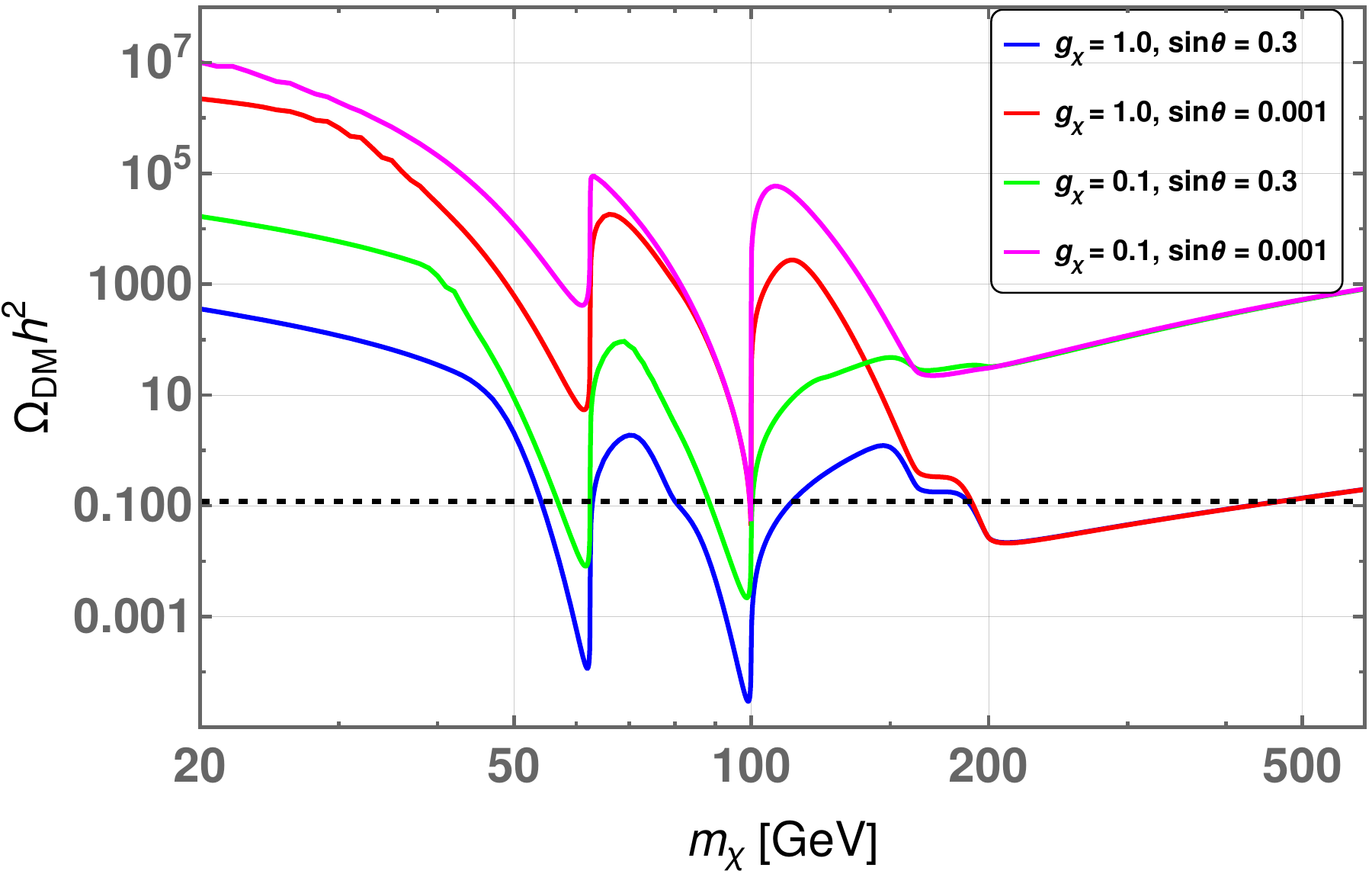}
    \includegraphics[width=0.49\linewidth]{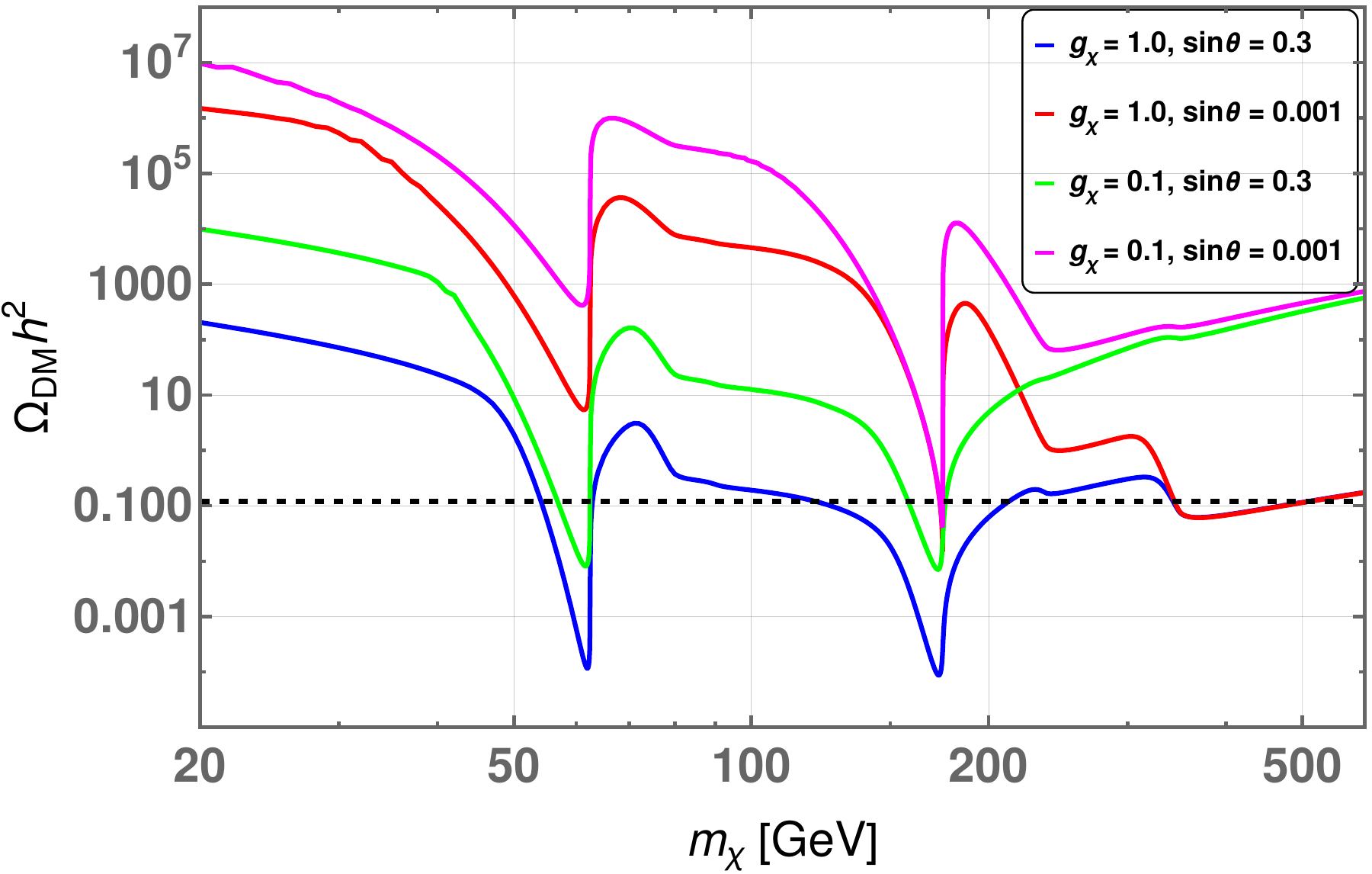}
    \caption{Variation of relic abundance with dark matter mass $m_\chi$ with two different chosen values of $\sin \theta$ and $g_{\chi}$ mentioned in the inset. Left panel for $m_{h_2}=200$ GeV and right for $m_{h_2}=350$ GeV. For both cases, we have taken $\lambda_{hs}=0.8$ and $\mu_3 = -50$ GeV. The black dashed line indicates the observed dark matter relic abundance,
$\Omega_{\rm DM} h^2 = 0.12 \pm 0.001$~\cite{Planck:2018vyg}.
}
    \label{fig:relic_mass}
\end{figure}

The interplay between $\sin\theta$ and $g_\chi$ yields the following behavior, which has been displayed in 
Fig.\,\ref{fig:relic_mass}. This can be summarized as follows:

\begin{itemize}

\item \textbf{Both $\sin\theta$ and $g_\chi$ large:}  
The annihilation cross section receives substantial enhancement from both scalar 
mediators. In this case, the observed relic density can be reproduced in \emph{all} the 
characteristic regions: the Higgs funnel, the $h_2$-resonance, and the degenerate region. The blue curve in Fig.\,\ref{fig:relic_mass} shows this region.

\item \textbf{Large $\sin\theta$, small $g_\chi$:}  
A sizeable mixing angle ensures efficient annihilation through the SM channels mediated 
by $h_1$ and $h_2$. The correct relic density is typically obtained in the Higgs funnel 
and the $h_2$-resonance regions. However, the degenerate region becomes ineffective because a small $g_\chi$ cannot 
generate sufficient annihilation away from the resonance peaks. The green line in Fig.\,\ref{fig:relic_mass} corresponds to this case. 

\item \textbf{Small $\sin\theta$, large $g_\chi$:}  
In this regime, the SM-mediated channels are suppressed, but the direct coupling to the 
singlet mediator is strong. The observed relic abundance can then be obtained in the 
$h_2$-resonance region and in the degenerate domain, where processes such as 
$\chi\chi\rightarrow h_2^\ast \to \text{SM SM}$ or $\chi\chi\rightarrow h_2 h_2$ may dominate. In Fig.\,\ref{fig:relic_mass}, the red curve highlights this region.

\item \textbf{Both $\sin\theta$ and $g_\chi$ small:}  
Here, the annihilation cross section remains insufficient throughout most of the 
parameter space. The correct relic density can be achieved \emph{only} near a very narrow region of the
$h_2$-resonance, shown in magenta contour in Fig.\,\ref{fig:relic_mass}, where the s-channel enhancement boosts the annihilation efficiency.

\end{itemize}

\begin{figure}[h!]
    \centering
    \includegraphics[width=0.49\textwidth]{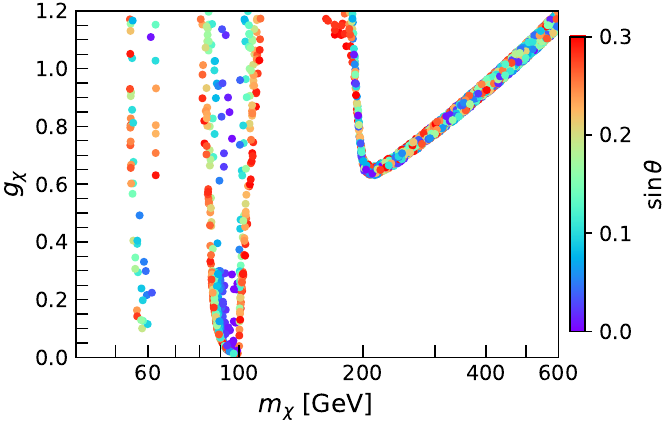}
    \includegraphics[width=0.49\textwidth]{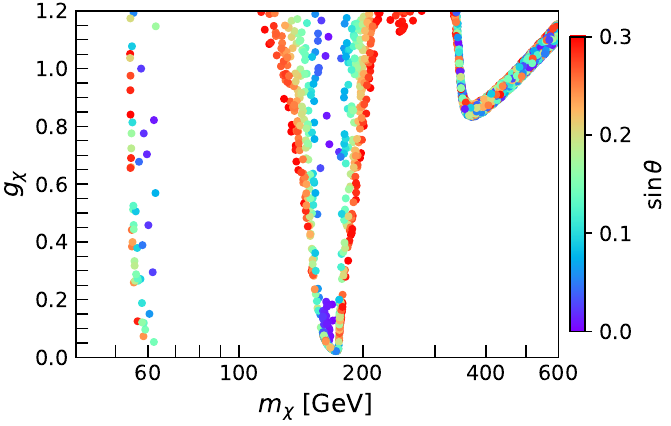}
    \caption{
        Variation of the Yukawa coupling \( g_\chi \) with DM mass \( m_\chi \) maintaining relic abundance in the range \( 0.1 \le \Omega_{\rm DM} h^2 \le 0.121 \). The color gradient represents the mixing angle \( \sin\theta \). Left panel for $m_{h_2}=200$ GeV and right for $m_{h_2}=350$ GeV. For both cases, we have taken $\lambda_{hs}=0.8$ and $\mu_3 = -50$ GeV.}
    \label{fig:yukawa}
\end{figure}
Overall, these patterns clearly indicate that achieving the observed thermal relic density 
requires the DM to be annihilated either through a sufficiently large mixing angle in the scalar sector
or through a sizeable DM Yukawa interaction. We have displayed the relic abundance against the DM mass in Fig.\,\ref{fig:relic_mass} for some representative fixed values of the mixing angle $\sin \theta$ and coupling constant $g_{\chi}$. The purpose is to highlight the dependency of the relic abundance on the mass of the DM. One may take, in principle, any mixing angle and coupling constant in between the representative points, but the observed relic density can be obtained only in the three regions of the DM mass as mentioned above.

The thermal relic abundance in this scalar-portal framework is governed by the interplay between the scalar mixing angle $\sin\theta$ and the dark-sector Yukawa coupling $g_\chi$. The left panel of Fig.\,\ref{fig:yukawa} shows the allowed parameter space in the $(m_\chi,\, g_\chi)$ plane, where the color gradient encodes the value of the mixing angle $\sin\theta$, for a fixed heavy-scalar mass $m_{h_2}=200~\text{GeV}$, portal coupling $\lambda_{hs}=0.8$, and trilinear parameter $\mu_3 = -50~\text{GeV}$. The displayed region corresponds to parameter values that yield a relic density within the %observationally allowed 
range, $0.1 \leq \Omega_{\rm DM} h^2 \leq 0.121$ \cite{Planck:2018vyg}.

As seen from this figure, the observed relic abundance can be reproduced across a broad interval of $g_\chi$ values in both the Higgs-funnel region and in the vicinity of the heavy-scalar ($h_2$) resonance. In the Higgs-funnel regime ($m_\chi \simeq m_{h_1}/2$), the annihilation cross section is primarily controlled by the Higgs--portal interaction, which scales with the mixing angle. Consequently, obtaining the correct relic density in this region requires a relatively large $\sin\theta$, while the precise value of $g_\chi$ plays only a subdominant role.

In contrast, when the dark-matter mass approaches the heavy-scalar resonance, $m_\chi \simeq m_{h_2}/2$, the Breit-Wigner enhancement substantially increases the annihilation rate. This resonant behavior permits a wide range of mixing angles to satisfy the relic density constraint, and largely independent of the magnitude of $g_\chi$ (depending on how close the DM mass is to $m_{h_2}/2$). However, away from this near-degenerate region, the annihilation cross section becomes insufficient for small $g_\chi$, and even large values of $\sin\theta$ fail to reproduce the observed relic abundance. The right panel of Fig.\,\ref{fig:yukawa}, corresponding to a heavier scalar mass ($m_{h_2}=350~\text{GeV}$), is shown for the same trilinear and Higgs portal couplings as in the left panel. The qualitative behavior remains similar to that observed in the left panel; however, the viable parameter space is shifted toward larger dark-matter masses, reflecting the dependence of the annihilation dynamics on the heavy scalar mass scale.

\begin{figure}[h!]
    \centering
    \includegraphics[width=0.5\textwidth]{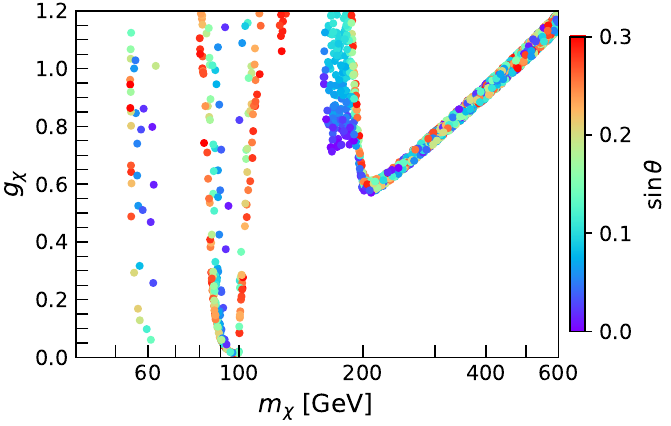}
    \caption{
       Variation of \(g_\chi\) as a function of the dark matter mass \(m_\chi\) for a sizable Higgs portal coupling $\lambda_{hs}=2.0$, while all other parameters are fixed to the same values as in the left panel of Fig.\,\ref{fig:yukawa}.}
    \label{fig:yukawaH}
\end{figure}

The requirement of a strong first-order electroweak phase transition, discussed in a later section, favors relatively large values of the portal coupling $\lambda_{hs}$ for a larger singlet scalar mass. Fig.\,\ref{fig:yukawaH} shows that the observed relic density can be accommodated even for such enhanced values of $\lambda_{hs}$. In particular, a larger portal coupling allows for a broader region of viable parameter space in the regime $m_\chi \lesssim m_{h_2}$. Moreover, as discussed earlier, variations in $\lambda_{hs}$ do not significantly impact the determination of the direct detection cross section.

\begin{figure}
    \centering
    \includegraphics[width=0.49\linewidth]{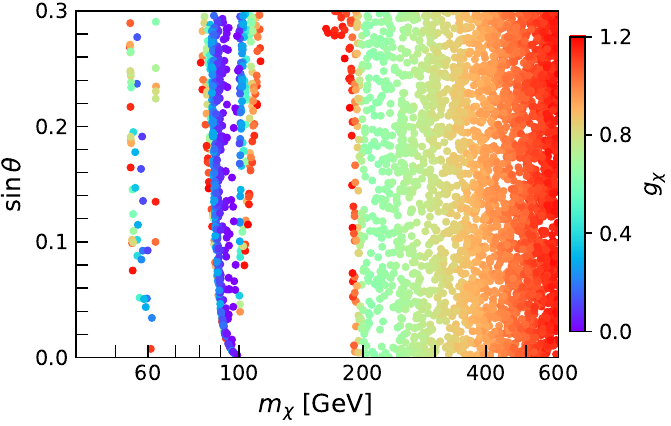}
    \includegraphics[width=0.49\linewidth]{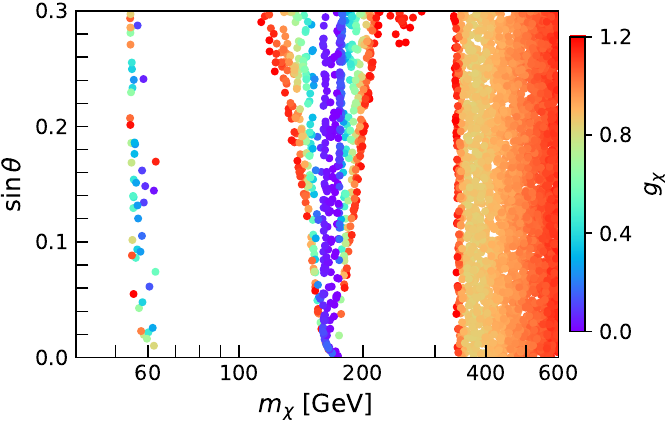}
    \caption{Variation of mixing angle $\sin\theta$ with DM mass $m_\chi$ maintaining relic abundance in the range $0.1\le\Omega_{\rm DM} h^2\le 0.121$. The color gradient represents the Yukawa coupling $g_\chi$. Left panel for $m_{h_2}=200$ GeV and right for $m_{h_2}=350$ GeV. For both cases, we have taken $\lambda_{hs}=0.8$ and $\mu_3 = -50$ GeV.}
    \label{fig:stheta}
\end{figure}

Complementary insight is provided by Fig.\,\ref{fig:stheta}, which shows the allowed parameter space in the $(m_\chi,\, \sin\theta)$ plane, with the color gradient indicating the corresponding values of the Yukawa coupling $g_\chi$. All other model parameters are fixed to the same values used in Fig.~\ref{fig:yukawa}. The scatter points represent parameter sets that reproduce the observed thermal relic abundance, $0.1 \leq \Omega_{\rm DM} h^2 \leq 0.121$ \cite{Planck:2018vyg}. As the heavy scalar mass increases, the annihilation cross section becomes progressively more sensitive to the dark-sector Yukawa interaction and less dependent on the scalar mixing angle. In this large-$m_{h_2}$ regime, the relic density constraint is therefore only weakly correlated with $\sin\theta$, and satisfying the required annihilation rate necessitates a sufficiently large $g_\chi$. This behaviour clearly demonstrates the transition from mixing-driven annihilation in the Higgs-funnel region to Yukawa-dominated annihilation as the heavy scalar mass scale is increased.

We have selected nine benchmark points (BPs) shown in Tab.\,\ref{tab:BPs} from various parts of the parameter space. The specified BPs remain viable under current experimental scrutiny, adhering to both cosmological relic density measurements and direct detection constraints. We perform our analysis primarily for two representative BSM scalar masses, $m_{h_2} = 200$ GeV and $350~\text{GeV}$. These mass choices ensure that the additional scalar remains within the LHC kinematic reach, allowing for potential production and providing an immediate experimental test of the model. For one of our benchmark scenarios (BP9), we consider \( m_{h_2} \) to be less than the mass of the SM Higgs boson. This choice makes the selected set of benchmark points\footnote{To achieve a SFOEWPT, several benchmarks require large quartic couplings ($\lambda_{hs}$ and $\lambda_{s}$). These values may trigger Landau poles at an intermediate scale, rendering the model as an effective field theory valid only up to a cutoff $\Lambda$. However, because this study targets electroweak-scale phenomena—such as relic density, direct detection, collider observables, and gravitational wave signals—this feature does not impact our conclusions.} exhaustive, thereby rendering our systematic investigation complete, as it covers a wide region of parameter space that satisfies various collider bounds while simultaneously reproducing the correct DM and  EWPT observables. 

We do not explore values of $m_{h_2}$ significantly above $350~\text{GeV}$, since heavier singlet-like state tend to decouple from the Higgs sector. This decoupling makes it increasingly difficult to achieve a SFOEWPT which will be discussed in the next section. Compensating for this behavior would require taking the portal coupling $\lambda_{hs}$ to very large values, which is theoretically disfavored due to perturbativity constraints. For these reasons, our analysis is restricted to the region $m_{h_2} \lesssim 350~\text{GeV}$.

Achieving a SFOEWPT usually requires an %$\mathcal{O}(1)$ 
large Higgs portal coupling, which, in turn, inevitably opens the $h_1 \to h_2 h_2$ decay mode in this mass range, adding a sizable partial width to the Higgs boson total width. Latest searches for invisible decays and measurement of total width of the Higgs boson at the LHC put stringent constraints for such mass range $m_{h_2} < m_{h_1}/2$ \cite{ATLAS:2023tkt}. Therefore, we decide to leave aside the region $m_{h_2} < m_{h_1}/2$ in our analysis.

%%%%%%%%%%%%%%%%%%%%%%%%%%%%
%%%%%%%%%%%%%%%%%%%%%%%%%%

\begin{table}[ht]
    \centering
    \begin{tabular}{|c|c|c|c|c|c|c|c|}
    \hline
       BPs & $m_{h_2}$\small [GeV] & $\sin\theta$  &  $\lambda_{hs}$ & $\lambda_{s}$& $\mu_3$\small[GeV] & $m_\chi$\small [GeV] & $g_\chi$ \\
%       & & &  &  & & & \\
         \hline
        BP1  &  &   & 2.2 & 3.38 & -20.0 & 222.0  & 0.57\\
        % & & &&&&&\\
        \cline{4-8}
        BP2 & 200.0 & $0.001$ & 2.4 & 3.1 & 100.0 & 242.0  & 0.56 \\
        % & & &&&&&\\
        \cline{4-8}
       BP3 & &   & 1.9  & 2.95 & -100.0 & 180.0  & 0.76 \\
        % & & &&&&&\\
        \hline
        BP4   &  &   & 4.3 & 4.5 & 20.0 & 310.0  & 0.87 \\
        % & & &&&&&\\
       \cline{4-8}
        BP5 & 350.0 & $0.001$ & 4.45 & 4.0  & 100.0 & 380.0  & 0.73 \\
      %   & & &&&&&\\
        \cline{4-8}
        BP6 &  &   & 4.0  & 4.1 & -100.0 & 300.0  & 0.87 \\
       %  & & &&&&&\\
        \hline
        BP7  & 350.0 & 0.07 & 3.8 & 4.8 & 20.0 &  167.4 & 0.03 \\
%         & & &&&&&\\
       \hline
        BP8 & 200.0 & $0.13$ & 2.02 & 3.0  & 20.0 & 94.75  & 0.02 \\
        \hline
      BP9 & 70.0 & $0.002$ & 0.8 & 0.73  & 20.0 & 78.0  & 0.34 \\
        \hline
    \end{tabular}
    \caption{Benchmark points for singlet scalar extension of the SM along with singlet fermion dark matter. All the BPs satisfy the observed relic density and bound from the upper limit of spin-independent DM-nucleon scattering cross section.}
    \label{tab:BPs}
    \end{table}

Tab.\,\ref{tab:dd} presents the corresponding dark matter observables for the same benchmark points, including the relic abundance ($\Omega_{\rm DM}h^2$) and the thermally averaged annihilation cross sections. An interesting observation is that the DM predominantly annihilate into $h_1 h_1, h_2 h_2, h_1 h_2$ final states in most of the BPs. The dearth of SM fermionic channels in the DM annihilation is due to the presence of small $\sin\theta$ and $m_q$ in both the vertices of the Feynman diagram. Moreover, all the annihilation to SM processes are kinematically s-channel suppressed except at the Higgs resonance. In BP7 and BP8 with relatively larger $\sin \theta$, DM annihilates to SM gauge bosons channels. In most of the BPs the dominant annihilation channels are the di-scalar boson final states. Though there may appear $\sin \theta$ in the triple scalar couplings, but those couplings, at the same time, are proportional to $\lambda_{hs}v$. Here, large $\lambda_{hs}$ in the BPs would play a significant role. In particular, there is no $\sin \theta$ factor in the $h_1-h_2-h_2$ coupling. Therefore, a large fraction of the annihilation process $\chi \chi \to h^{*}_2 \to h_1 h_2$ (in BP4,6) can be observed where both vertices are driven by $\cos \theta$. Similarly, in BP1 and BP2, DM mostly annihilates into $h_2 h_2$ via predominantly $\cos^2 \theta$ enhanced t-channel process.  In addition to this, the dimensionful parameter $\mu_{hs}$ will appear in $h_2-h_1-h_1$ coupling, which fuels $\chi \chi \to h_2 \to h_1 h_1$ channel (in BP3).

\subsection{Direct detection}

In this model, the singlet fermion DM $\chi$ interacts with the nuclei through $t$-channel 
exchange of the two CP-even mass eigenstates $h_1$ and $h_2$. Before EWSB, only the SM Higgs doublet $H$ couples to quarks, while the singlet real scalar field $s$ couples to the dark matter fermion $\chi$ through Yukawa interaction. EWSB enables mixing between the scalars, which introduces the interaction between the DM and light quarks inside the nuclei. After EWSB, the couplings of the two scalars with the mass eigenstates to quarks and to dark matter become:
\begin{align}
g_{\chi\chi h_1} &= -g_\chi \sin\theta, 
& g_{qqh_1} &= \frac{m_q}{v}\cos\theta, \\[4pt]
g_{\chi\chi h_2} &= \;\,g_\chi \cos\theta, 
& g_{qqh_2} &= \frac{m_q}{v}\sin\theta .
\end{align}
At the momentum transfers relevant for direct detection, both scalars can be 
integrated out, which yields the effective scalar interaction
\begin{equation}
\label{eq:Leff-DD}
\mathcal{L}_{\rm eff}
=
\sum_q
\left[
g_\chi\,\frac{m_q}{v}\,\sin\theta\cos\theta
\left(
\frac{1}{m_{h_2}^2} - \frac{1}{m_{h_1}^2}
\right)
\right]
(\bar\chi\chi)\,(\bar q q).
\end{equation}
This form exhibits the characteristic destructive interference between $h_1$ and 
$h_2$ exchange: the opposite signs of the effective coefficients arise directly 
from the single mixing angle. Matching Eq.~\eqref{eq:Leff-DD} to the nucleon scalar 
matrix element,
\begin{equation}
\langle N | \sum_q m_q \bar q q | N\rangle = f_N m_N \bar N N,
\qquad f_N \simeq 0.30,
\end{equation}
leads to the familiar spin-independent (SI) DM--nucleon scattering cross section,
\begin{equation}
\sigma_{\rm SI}
=
\frac{f_N^2 m_N^4}{\pi (m_\chi + m_N)^2}
\left[
\frac{g_\chi\,\sin\theta\cos\theta}{v}
\left(
\frac{1}{m_{h_2}^2} - \frac{1}{m_{h_1}^2}
\right)
\right]^2 .
\label{eq:sigmaSI-final}
\end{equation}
Fig.\,\ref{fig:dd} shows the predicted SI cross section against DM mass for various $(\sin\theta$ compared with the latest limits from the LZ (2024) experiment~\cite{LZ:2024zvo}. For this figure, we consider only the spin-independent cross section for the neutron, which exceeds that for the proton and thus provides a conservative estimate. The spin-independent DM-proton/neutron scattering cross sections ($\sigma_{\rm SI}$) for all the BPs are also shown in table~\ref{tab:dd}. Several qualitative features follow directly from Eq.~\eqref{eq:sigmaSI-final}:

\begin{figure}[h!]
    \centering
    \includegraphics[width=0.49\linewidth]{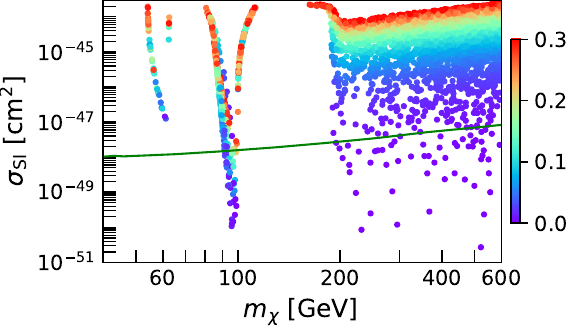}
    \includegraphics[width=0.49\linewidth]{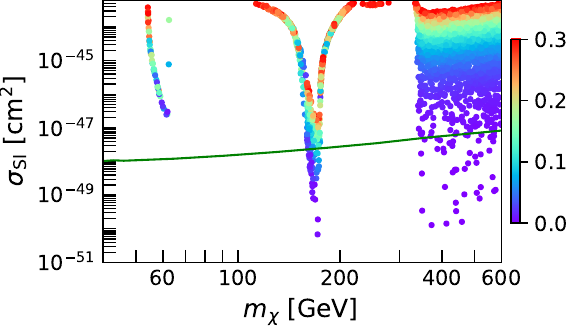}
    \caption{Variation of nucleon-dark matter spin independent scattering cross-section $\sigma_{\rm SI}$ with DM mass $m_\chi$ maintaining relic abundance in the range $0.1\le\Omega_{\rm DM} h^2\le 0.121$. The color gradient represents the mixing angle $\sin\theta$. Left panel for $m_{h_2}=200$ GeV and right for $m_{h_2}=350$ GeV. For both cases, we have taken $\lambda_{hs}=0.8$ and $\mu_3 = -50$ GeV. The solid green line represents the latest $2\sigma$ bound from the LZ (2024) experiment~\cite{LZ:2024zvo}.}
    \label{fig:dd}
\end{figure}

\begin{table}[httb]
\centering
\renewcommand{\arraystretch}{1.3}
\begin{tabular}{|c|c|c|c|c|}
\hline
BPs
& $\Omega_{\rm DM} h^2$ 
& $\sigma_{\rm SI}^{(p)}~(\mathrm{cm}^2)$ 
& $\sigma_{\rm SI}^{(n)}~(\mathrm{cm}^2)$ 
& \shortstack{Annihilation\\channels} \\
\hline

BP1
& 0.119   
& $6.96\times 10^{-50}$ 
& $7.10\times 10^{-50}$ 
& \shortstack{$h_2 h_2~(64\%)$\\$h_1 h_1~(36\%)$} \\
\hline

BP2
& 0.119   
& $6.72\times 10^{-50}$ 
& $6.86\times 10^{-50}$ 
& \shortstack{$h_2 h_2~(71\%)$\\$h_1 h_1~(29\%)$} \\
\hline

BP3
& 0.119   
& $1.24\times 10^{-49}$ 
& $1.26\times 10^{-49}$ 
& \shortstack{$h_1 h_1~(99\%)$\\$h_2 h_2~(1\%)$} \\
\hline

BP4
& 0.120   
& $3.26\times 10^{-49}$ 
& $3.32\times 10^{-49}$ 
& \shortstack{$h_1 h_2~(100\%)$} \\
\hline

BP5
& 0.120   
& $2.35\times 10^{-49}$ 
& $2.40\times 10^{-49}$ 
& \shortstack{$h_2 h_2~(69\%)$\\$h_1 h_1~(31\%)$} \\
\hline

BP6
& 0.120 
& $3.33\times 10^{-49}$ 
& $3.40\times 10^{-49}$ 
& \shortstack{$h_1 h_2~(100\%)$} \\
\hline

BP7
& 0.119 
& $1.92\times 10^{-48}$ 
& $1.96\times 10^{-48}$ 
& \shortstack{$h_1 h_1~(67\%)$\\$W^+W^-~(23\%)$\\$ZZ~(7\%)$} \\
\hline

BP8
& 0.120 
& $1.40\times 10^{-48}$ 
& $1.44\times 10^{-48}$ 
& \shortstack{$W^+W^-~(74\%)$\\$ZZ~(26\%)$} \\
\hline

BP9
& 0.119 
& $1.26\times 10^{-48}$ 
& $1.28\times 10^{-48}$ 
& \shortstack{$h_2 h_2~(100\%)$} \\
\hline

\end{tabular}
\caption{Dark matter observables for the benchmark points, including relic abundance, spin-independent scattering cross sections off protons and neutrons, and dominant annihilation channels.}
\label{tab:dd}
\end{table}

\begin{itemize}

\item Large scalar mixing angle yields a large DM-nucleon scattering cross section, making large $\sin \theta$ untenable against the latest LZ (2024) results~\cite{LZ:2024zvo}.  

\item Only for small $\sin\theta$, the mediator--nucleon coupling becomes 
suppressed, pushing the predicted cross section below the strongest upper limit of spin-independent DM-nucleon scattering cross section given by the LZ experiment. In this regime, sizeable $g_\chi$ can still yield detectable rates through $h_2$ 
exchange, provided that $m_{h_2}$ is not too large.

\item The expression inside bracket in Eq.~\eqref{eq:sigmaSI-final} can become very small when the 
two propagators nearly cancel.  This ``destructive interference'' or ``blind-spot'' behaviour produces the 
dip visible in Fig.~6 for certain combinations of $(m_{h_2}, \sin\theta)$
and allows otherwise large-coupling scenarios to evade present bounds.

\end{itemize}
Overall, direct detection searches significantly restrict regions of parameter space that also produce the correct thermal relic density.  In particular, the relic-favored regions with either large $\sin\theta$ or large $g_\chi$ are often in tension with the XENON1T and LZ bounds, whereas the $h_2$-resonance region remains largely viable due to suppressed couplings and the interference effect described above.

Therefore, the key points extracted from the analysis of relic density and direct detection over the BPs could be summarized as:
\begin{itemize}
 \item At the resonance of $h_2$ (\textit{i.e.,} when $m_\chi \simeq m_{h_2}/2$), the dark-matter relic density becomes essentially independent of all parameters except the dark-matter mass $m_\chi$, which determines the proximity to the $h_2$ resonance, and the Yukawa coupling $g_\chi$, which controls the annihilation rate. Since the relic density is insensitive to $\sin\theta$ in this regime, the mixing angle may be taken up to the maximum value allowed by direct-detection bounds as considered in BP7 and BP8 in Tab.\,\ref{tab:BPs}.
 \item Throughout the entire parameter space, the relic density is independent of the scalar quartic coupling $\lambda_s$. This coupling does not affect the dominant dark-matter annihilation processes and therefore does not influence the relic-abundance calculation.
 \item In the region where the dark-matter mass satisfies $m_\chi \gtrapprox m_{h_2}$, the relic density is mostly insensitive to $\sin \theta$, but can be achieved only for relatively large values of $g_\chi$ (e.g., BP1, BP2 BP5 and BP9). However, larger $g_{\chi}$ enhances spin independent direct detection cross section. Therefore, $\sigma_{\rm SI}$ permits only very small values of the mixing angle $\sin\theta$. 
\item In the Higgs-funnel region, where $m_{\chi} \approx m_{h_1}/2$ and thus lies far from the degenerate regime $m_{\chi} \approx m_{h_2}$, the correct dark-matter relic density can be obtained only for comparatively large values of the mixing angle $\sin\theta$. However, such large mixing angles are generally excluded by current direct-detection constraints.

\item For BP1 to BP6, where $m_{\chi} \simeq m_{h_2}$ (the degenerate region), a relatively large coupling $g_{\chi}$ is required to achieve the correct relic abundance. In contrast, in BP9 the relic density can be reproduced with a smaller value of $g_{\chi}$, owing to the lighter mass of the scalar $h_2$.

\item Note that if the VEV of the singlet scalar is temperature dependent, the mass of the DM would also be temperature dependent, which will, in principle, modify the thermal freeze-out. However, in our setup, dark matter freeze-out occurs well within the electroweak broken phase, after the scalar VEVs have settled close to their zero-temperature values. Therefore, the residual thermal shifts in the relevant masses and mixing parameters are negligible, and their effects on the annihilation rate and final relic abundance are negligible as well. 
\end{itemize}

\subsection{Future collider prospects}
The non-resonant production of \(h_{1}h_{2}\) and \(h_{2}h_{2}\) at the LHC serves as a direct probe of the extended Higgs potential. While these processes are suppressed by the small mixing angle \(\sin \theta \), their cross sections are fundamentally governed by the trilinear couplings \(\lambda _{112}\) and \(\lambda _{122}\). These couplings can remain substantial even when mixing is minimal, reflecting the underlying symmetry-breaking scale of the singlet sector.

We compute the partonic production cross sections for the non-standard di-scalar
final states \(pp\to h_1 h_2\) and \(pp\to h_2 h_2\) at \(\sqrt{s}=14~\mathrm{TeV}\).
The model is implemented in \texttt{FeynRules}~v2.3~\cite{Christensen:2008py},
with the UFO output generated using \texttt{NLOCT}~\cite{Degrande:2014vpa} and
the UFO interface~\cite{Degrande:2011ua,deAquino:2011ub}. The cross sections are
then evaluated in \texttt{MadGraph5\_aMC@NLO}~\cite{Alwall:2011uj} using the
NNPDF23LO PDFs~\cite{Ball:2013hta,NNPDF:2014otw}. 

For BP1-BP7, the predicted cross sections for both \(pp\to h_1 h_2\) and \(pp\to h_2 h_2\) remain far below the femtobarn level, reflecting the suppressed production of the non-SM scalar states induced by the small Higgs-singlet mixing.
An exception is BP8, where the di-scalar rates are comparatively enhanced and reach
the sub-femtobarn level, making this benchmark the most promising target among the
set for HL-LHC or future hadron-collider searches. While BP9 features a much lighter
\(h_2\), in this benchmark, the non-standard di-scalar signal is driven primarily by the available phase space rather than mixing enhancement, and its detailed collider prospects depend on dedicated analyses beyond the scope of the present work.

An interesting observation is that the enhancement observed for BP8 is correlated with a resonant configuration in which the dark-matter mass lies
close to $m_{h_2}/2$. To verify this behaviour, we explicitly vary $m_\chi$ in the vicinity of the resonance and find that the scalar mixing angle can increase to moderately larger values while remaining consistent with current collider and direct-detection constraints. As a consequence, the associated production rate of the heavier scalar is significantly enhanced. In this resonant regime, the associated production of the heavier scalar can approach the femtobarn level, potentially placing such scenarios within the projected sensitivity of the HL-LHC and making them interesting targets for future high-energy hadron colliders, such as the FCC-hh. A detailed collider-level analysis of these channels is deferred to future work, as the present study
primarily focuses on cosmological probes, in particular the electroweak phase transition and the associated stochastic gravitational-wave signatures.

While direct-detection experiments and collider searches impose strong and
complementary constraints on the parameter space, they do not fully probe
the regions favoured by dark-matter phenomenology. In particular, the
requirement of satisfying current direct-detection bounds forces the scalar
mixing angle to be very small, which in turn strongly suppresses the
production rates of the heavy scalar at colliders. As a result, much of the
viable parameter space is simultaneously difficult to access through both
direct-detection experiments and collider searches.

This motivates the exploration of complementary probes that are sensitive to
the thermal history of the scalar sector. In particular, the dynamics of the EWPT provide a powerful handle on the underlying
scalar interactions. A SFOEWPT sources a stochastic GW background, offering an alternative
observational window into regions of parameter space that may remain
inaccessible to conventional collider and direct-detection probes. We
therefore proceed to analyse the EWPT and the
associated GW signatures of the model.

%%%%%%%%%%%%%%%%%
%%%%%%%%%%%%%%%%
\section{Electroweak phase transition and GW detection}\label{sec:ewpt}
To study the EWPT, we analyze the finite-temperature effective potential, starting from the one-loop zero-temperature potential and incorporating the thermal corrections that shape the phase transition dynamics. We then examine the associated stochastic GW signal and assess its detectability in future experiments.

\subsection{One-loop effective potential}
The one-loop effective potential depends entirely on the field-dependent mass-squared 
eigenvalues $m_i^2(\varphi)$ of all bosonic and fermionic degrees of freedom, obtained by diagonalizing the corresponding tree-level mass matrices evaluated on the background 
field configuration $\varphi=(h,s)$. For the CP-even scalar sector, the field-dependent 
mass-squared matrix elements are
\begin{align}
m_{hh}^2(h,s) &\equiv \frac{\partial^2 V_0}{\partial h^2}
= \mu_h^2 + 3\lambda_h h^2 + \frac{\lambda_{hs}}{2}s^2 + \mu_{hs}s,\label{eq:mat1} \\
m_{ss}^2(h,s) &\equiv \frac{\partial^2 V_0}{\partial s^2}
= \mu_s^2 + 3\lambda_s s^2 + \frac{\lambda_{hs}}{2}h^2 + 2\mu_3 s, \\
m_{hs}^2(h,s) &\equiv \frac{\partial^2 V_0}{\partial h\,\partial s}
= \lambda_{hs}hs + \mu_{hs}h\label{eq:mat2} .
\end{align}
For the CP-odd Goldstone bosons $\chi_i$, the field-dependent mass-squared is
\begin{eqnarray}
m_{\chi_i}^2(h,s)
= \mu_h^2 + \lambda_h h^2 + \frac{\lambda_{hs}}{2}s^2 + \mu_{hs}s .
\end{eqnarray}
The field-dependent masses of the SM fermions and electroweak gauge bosons read
\begin{eqnarray}
m_i^2(h) = \frac{1}{2}y_i^2 h^2, \qquad 
m_W^2(h)=\frac{g^2}{4}h^2, \qquad 
m_Z^2(h)=\frac{g^2+g^{\prime 2}}{4}h^2 ,
\end{eqnarray}
where \( y_i = \sqrt{2}\, m_i / v \) denotes the Yukawa coupling of fermion species \( i \), where \( m_i \) is the corresponding fermion mass at zero temperature. The parameters \( g \) and \( g' \) are the gauge couplings associated with the \( SU(2)_L \) and \( U(1)_Y \) gauge groups. For the numerical analysis, we include all charged leptons and quarks.

The field-dependent mass of the BSM Dirac fermion $\chi$ is
\begin{eqnarray}
m_\chi(s) = m_\chi - g_\chi\, s ,
\end{eqnarray}
where $g_\chi$ is the Yukawa coupling mentioned earlier.

\begin{comment}
\end{comment}

The one-loop correction to the zero-temperature effective potential, known as the Coleman–Weinberg (CW) potential~\cite{Coleman:1973jx}, arises from the sum of all one-particle-irreducible vacuum diagrams and encodes quantum fluctuations around the classical background field. In $\overline{MS}$ scheme, its explicit form follows the standard expression~\cite{Quiros:1999jp}
\begin{eqnarray}\label{eq:CWpot}
    V_{\rm 1-loop}^{\rm CW} (h,s)
= \frac{1}{64\pi^2} \sum_{i}n_i m_i^4(h,s) \left[ \log \left( \frac{m_i^2(h,s)}{\mu^2} \right) - C_i \right].
\end{eqnarray}
Here, $\mu$ denotes the renormalization scale of the theory. The constants $C_i$ take the value $3/2$ for scalars, fermions, and the longitudinal components\footnote{At high temperatures, the photon acquires field-dependent thermal mass and mixes with the longitudinal component of the \(Z\) boson, as discussed later.} of gauge bosons, while $C_i = 1/2$ for the transverse gauge modes~\cite{Wainwright:2011kj,Blinov:2015vma,Aoki:2021oez}. The corresponding degrees of freedom are assigned in the usual manner for bosonic and fermionic fields\footnote{A closed fermion loop contributes an extra minus sign due to the anticommutation of fermionic fields in the trace, which is absorbed into the corresponding degrees of freedom.} are
\begin{eqnarray}\label{eq:degrees}
n_{W_L} = 2,\, n_{W_T} = 4,\, n_{Z_L/\gamma_L} = 1,\, n_{Z_T} = 2,\,n_{h/s/\chi_i} = 1,\, n_q = -12,\, n_{\ell/\chi} = -4.
\end{eqnarray}
 Here, $q$ and $\ell$ denote the quarks and charged leptons, respectively. 
In the $\overline{\mathrm{MS}}$ renormalization scheme, the choice of the
renormalization scale $\mu$ introduces an intrinsic theoretical uncertainty
in thermodynamic quantities such as the critical and nucleation temperatures,
as well as in derived observables relevant for the electroweak phase
transition~\cite{Chiang:2018gsn,Athron:2022jyi,Croon:2020cgk,Gould:2021oba}. The explicit scale dependence of the one-loop
Coleman-Weinberg potential can, in principle, be reduced by employing an
RGE-improved treatment of the effective potential, as formulated in
Refs.~\cite{Andreassen:2014eha,Andreassen:2014gha}.

In this work, we do not attempt a systematic study of renormalization-scale or
gauge uncertainties. Our analysis is instead intended to identify regions of
parameter space that robustly exhibit a strong first-order electroweak phase
transition and potentially observable gravitational-wave signals. For the
benchmark points considered, we have checked that moderate variations of the
renormalization scale around the characteristic mass scales do not alter the
qualitative features of the phase transition.

More refined treatments based on dimensionally reduced effective field theory
approaches can further reduce theoretical uncertainties and are left for
future work~\cite{Gould:2021dzl,Gould:2023jbz,Chala:2024xll,Chala:2025oul,Navarrete:2025yxy,Biekotter:2025npc}.
For definiteness, throughout our numerical analysis, we fix the renormalization
scale to $\mu = m_t$.

The one-loop CW correction generically shifts both the vacuum location and the curvature of the scalar potential, thereby modifying the VEVs and physical masses at zero temperature. These effects are removed by introducing a counterterm potential $V_{\rm ct}(h,s)$ that cancels the loop-induced deviations and restores the tree-level vacuum structure. The counterterms are fixed by renormalization conditions that enforce the tree-level stationarity of the potential at the chosen VEVs and preserve its curvature, ensuring that the Higgs and singlet masses remain at their tree-level values. The counterterm potential takes the form~\cite{Chen:2017qcz}
\begin{eqnarray}\label{eq:ctpot}
    V_{\rm ct}(h,s)   &=& \frac{\delta\mu_h^2}{2} h^2 + \frac{\delta\lambda_h}{4} h^4  + \frac{\delta\mu_{s}^2}{2} s^2 +  \frac{\delta\mu_{hs}}{2}h^2 s +\delta a_1s.
\end{eqnarray}
with the coefficients (e.g.,\,$\delta \mu_h^2$) fixed by these renormalization conditions. The procedure for determining the counterterm coefficients, along with their expressions in terms of derivatives of the CW potential, is presented in App.~\ref{app:counter-term}.

The one-loop finite-temperature contribution to the effective potential is given by
\cite{Dolan:1973qd,Weinberg:1974hy}:
\begin{equation} \label{eq:finiteT}
V^{T}_{\rm 1-loop}(h,s,T)
= \frac{T^4}{2\pi^2}
\left[
\sum_{B} n_B\, J_B\!\left(\frac{m_B^2(h,s)}{T^2}\right)
+
\sum_{F} n_F\, J_F\!\left(\frac{m_F^2(h,s)}{T^2}\right)
\right],
\end{equation}
where $B = \{W_{L/T}, Z_{L/T}, \gamma_L, h, s, \chi_{1,2,3}\}$ includes all bosonic degrees 
of freedom, and $F$ contains the six quarks, three charged 
leptons, and the BSM Dirac fermion $\chi$. The thermal bosonic and fermionic functions are
\begin{eqnarray}
J_{B/F}\!\left(\frac{m^2(h,s)}{T^2}\right)
= \int_0^\infty dx\, x^2
\log\!\left(1 \mp e^{-\sqrt{x^2 + m^2(h,s)/T^2}}\right),
\end{eqnarray}
with the upper (lower) sign corresponding to bosons (fermions). For particles whose field--dependent masses satisfy $m^{2}(h,s) \gg T^{2}$, the thermal
functions $J_{B,F}$ become exponentially Boltzmann suppressed. As a result, such heavy states contribute negligibly to the finite temperature effective potential and thus have no significant impact on the dynamics of the electroweak phase transition.

In the high-temperature limit, $T^2 \gg m^2$, these functions admit the standard expansions
\cite{Quiros:1999jp}:
\begin{align} \label{eq:expn1}
J_B\!\left(\frac{m^2}{T^2}\right)
&= -\frac{\pi^4}{45}
+ \frac{\pi^2}{12}\frac{m^2}{T^2}
- \frac{\pi}{6}\frac{(m^2)^{3/2}}{T^3}
- \frac{m^4}{32T^4}
\ln\!\left(\frac{m^2}{a_b T^2}\right)
+ \cdots, \\
\label{eq:expn2}
J_F\!\left(\frac{m^2}{T^2}\right)
&= \frac{7\pi^4}{360}
- \frac{\pi^2}{24}\frac{m^2}{T^2}
- \frac{m^4}{32T^4}
\ln\!\left(\frac{m^2}{a_f T^2}\right)
+ \cdots,
\end{align}
where
\[
a_f = \pi^2 e^{3/2 - 2\gamma_E}, \qquad
a_b = 16\pi^2 e^{3/2 - 2\gamma_E},
\]
and $\gamma_E = 0.577$ is the Euler--Mascheroni constant. The first terms in the above two equations are field-independent and only shift the vacuum energy, leaving the phase transition dynamics unaffected. The dominant contributions at finite temperature arise from the quadratic $m^2$ terms; 
therefore, particles with lighter field-dependent masses contribute subleading corrections to the effective potential and have negligible impact on the phase transition.

 \subsection*{Ring resummation }
At finite temperature, bosonic contributions to the effective potential suffer from infrared divergences when the corresponding field-dependent masses become small~\cite{Senaha:2020mop,Kainulainen:2019kyp,Athron:2022jyi}. These divergences originate from the zero Matsubara modes of the bosonic degrees of freedom and are removed by incorporating the leading daisy (ring) diagrams. The ring-improved contribution to the effective potential is~\cite{Arnold:1992rz,Weinberg:1974hy,Carrington:1991hz}
\begin{equation}
V_{\rm ring}(h,s,T)
= -\frac{T}{12\pi}\, n_i 
\left[ \big(m_i^2(h,s,T)\big)^{3/2}
- \big(m_i^2(h,s)\big)^{3/2} \right],
\end{equation}
where \(m_i^2(h,s,T)\) are the thermally resummed Debye masses. These resummed masses are obtained by including the leading temperature-dependent self-energy corrections,
\begin{equation}
m_i^2(h,s,T) = m_i^2(h,s) + \Pi_i\, T^2,
\end{equation}
where \(\Pi_i\) represents the one-loop thermal self-energy of the corresponding field.

%\subsection*{Thermal Self-Energy Corrections}

Using the high-temperature expansion of the thermal functions, the CP-even scalar thermal self-energies are given by
\begin{align}
\Pi_{hh} &= \frac{3g^{2}}{16} + \frac{g'^{2}}{16}
+ \frac{y_q^{2}}{4} + \frac{y_\ell^{2}}{12}
+ \frac{\lambda_h}{2} + \frac{\lambda_{hs}}{24}, \\
\Pi_{ss} &=\label{eq:thFer} \frac{\lambda_s}{4} + \frac{\lambda_{hs}}{6}
+ \frac{g_\chi^{2}}{6}, \quad
\Pi_{hs} \approx 0.
\end{align}
As noted in~\cite{Blinov:2015vma,Ahriche:2007jp}, subleading thermal corrections to off-diagonal self-energies are suppressed by higher powers of couplings and electroweak VEV, and are therefore negligible. The CP-odd Goldstone bosons receive the same thermal correction as the Higgs direction,
\begin{equation}
\Pi_{\chi_i} = \frac{3g^{2}}{16} + \frac{g'^{2}}{16}
+ \frac{y_q^{2}}{4} + \frac{y_\ell^{2}}{12}
+ \frac{\lambda_h}{2} + \frac{\lambda_{hs}}{24}.
\end{equation}
As in the previously given expressions for the mass–squared matrix elements in Eqs.~\eqref{eq:mat1}--\eqref{eq:mat2}, once the thermal mass contributions 
(Debye masses) are included, the temperature-dependent mass-squared matrix of the CP-even scalars in the $(h,s)$ basis is given by
\begin{equation}
M^{2}(h,s,T)
=
\begin{pmatrix}
m_{hh}^2(h,s) & m_{hs}^2(h,s) \\
m_{hs}^2(h,s) & m_{ss}^2(h,s)
\end{pmatrix}
+
T^{2}
\begin{pmatrix}
\Pi_{hh} & 0 \\
0 & \Pi_{ss}
\end{pmatrix},
\end{equation}
and its eigenvalues correspond to the thermally corrected CP-even scalar masses. On the other hand, the longitudinal components of the electroweak gauge bosons acquire Debye masses in the thermal plasma. For the charged gauge bosons, the resummed thermal mass takes the form~\cite{Carrington:1991hz,Oikonomou:2024jms}
\begin{equation}
m_{W_L}^2(h,T)
= m_{W_L}^2(h) + \frac{11}{6} g^{2} T^{2}.
\end{equation}
In the neutral sector, the longitudinal components of the \(Z\) boson and the photon mix at finite temperature. The corresponding thermal mass-squared matrix is~\cite{Carrington:1991hz,Oikonomou:2024jms}
\begin{equation}
\mathcal{M}_L^{2}(h,T)
=
\begin{pmatrix}
\frac{1}{4} g^{2} h^{2} + \frac{11}{6} g^{2} T^{2} &
-\frac{1}{4} g g' h^{2} \\
-\frac{1}{4} g g' h^{2} &
\frac{1}{4} g'^{2} h^{2} + \frac{11}{6} g'^{2} T^{2}
\end{pmatrix},
\end{equation}
whose eigenvalues give the temperature and field-dependent masses of \(Z_L\) and \(\gamma_L\). 
%For $T=0$, one of the eigenvalues is zero, and it corresponds to the massless photon.

\begin{comment}
\end{comment}

\subsection{Electroweak phase transition}
The full one-loop effective potential at finite temperature is given by
\begin{eqnarray}\label{eq:effpot}
    V_{\rm eff}(h,s,T) &=& V_0(h,s) + V_{\rm 1-loop}^{\rm CW}(h,s,T) + V_{\rm 1-loop}^T(h,s,T) 
    + V_{\rm ct}(h,s)
\end{eqnarray}
where the individual contributions \(V_0(h,s)\), \(V^{\rm CW}_{1\text{-loop}}(h,s,T)\), \(V^{T}_{1\text{-loop}}(h,s,T)\), and \(V_{\rm ct}(h,s)\) are defined in
Eqs.~\eqref{eq:treepot}, \eqref{eq:CWpot}, \eqref{eq:finiteT}, and \eqref{eq:ctpot}, respectively.
The CW term \(V^{\rm CW}_{1\text{-loop}}(h,s,T)\) becomes explicitly temperature dependent once thermal screening is incorporated through Daisy (ring) resummation, which regulates infrared divergences from bosonic zero modes \cite{Senaha:2020mop,Kainulainen:2019kyp,Athron:2022jyi}. This may be done either by adding the ring contribution \(V_{\rm ring}(h,s,T)\)~\cite{Arnold:1992rz} or by replacing the bosonic masses with their thermally corrected counterparts \(m_i^2(h,T)\) \cite{Parwani:1991gq}; in this work, we follow the latter. Furthermore, each term in Eq.~\eqref{eq:effpot} acquires an implicit temperature dependence through the thermal evolution of the VEV of the scalars.

For a given choice of parameters, the critical temperature \(T_c\) and the corresponding
VEVs of the Higgs field \(v_c\) and the singlet field \(s_c\) are determined by the
conditions
\begin{eqnarray}
  &&\frac{\partial V_{\rm eff}(h,s,T_c)}{\partial h}\Big|_{\phi_{\rm low}} = 0,
  \qquad
    \frac{\partial V_{\rm eff}(h,s,T_c)}{\partial s}\Big|_{\phi_{\rm high}} = 0,\\
  && V_{\rm eff}(\phi_{\rm low},T_c) = V_{\rm eff}(\phi_{\rm high},T_c),
\end{eqnarray}
where the two degenerate vacua at \(T_c\) are given by
\(\phi_{\rm low} \equiv (h=v_c,\, s \approx 0)\) and
\(\phi_{\rm high} \equiv (h \approx 0,\, s=s_c)\).
The strength of the phase transition along each field direction is quantified by
\begin{eqnarray}
    \xi_h = \frac{v_c}{T_c}, \qquad 
    \xi_s = \frac{s_c}{T_c}.
\end{eqnarray}
A SFOEWPT requires \(\xi_h \gtrsim 1\) along the Higgs direction. In this work, we employ the \texttt{CosmoTransition} package~\cite{Wainwright:2011kj}
to determine the phase transition pattern and to compute \(T_c\) together with the field
VEVs at the critical temperature.

\begin{figure}\
    \centering
    \includegraphics[width=0.49\linewidth]{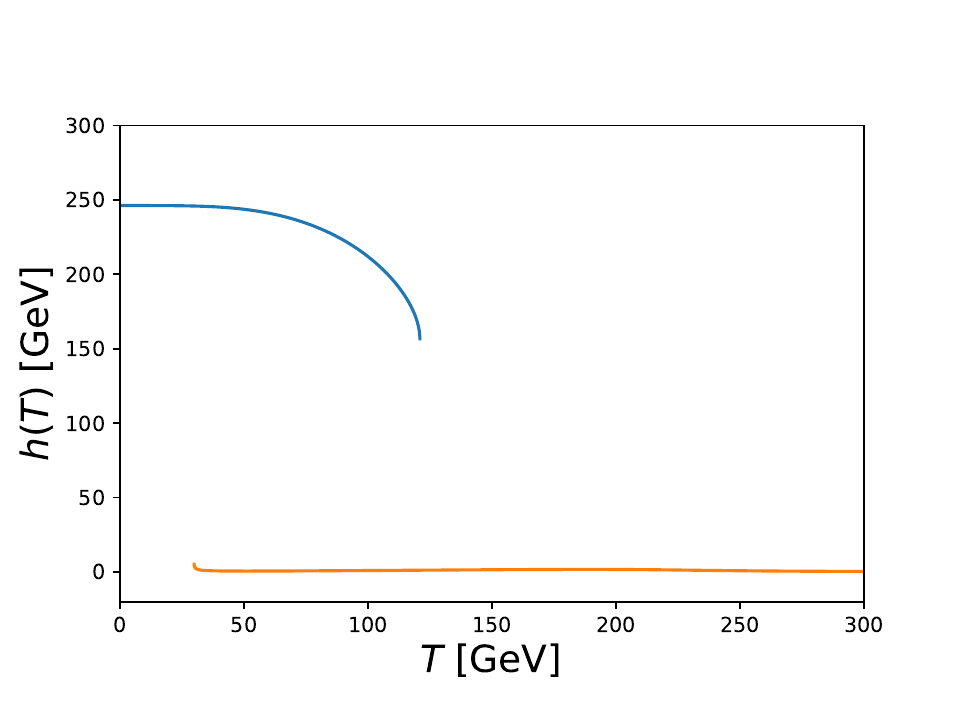}
    \includegraphics[width=0.49\linewidth]{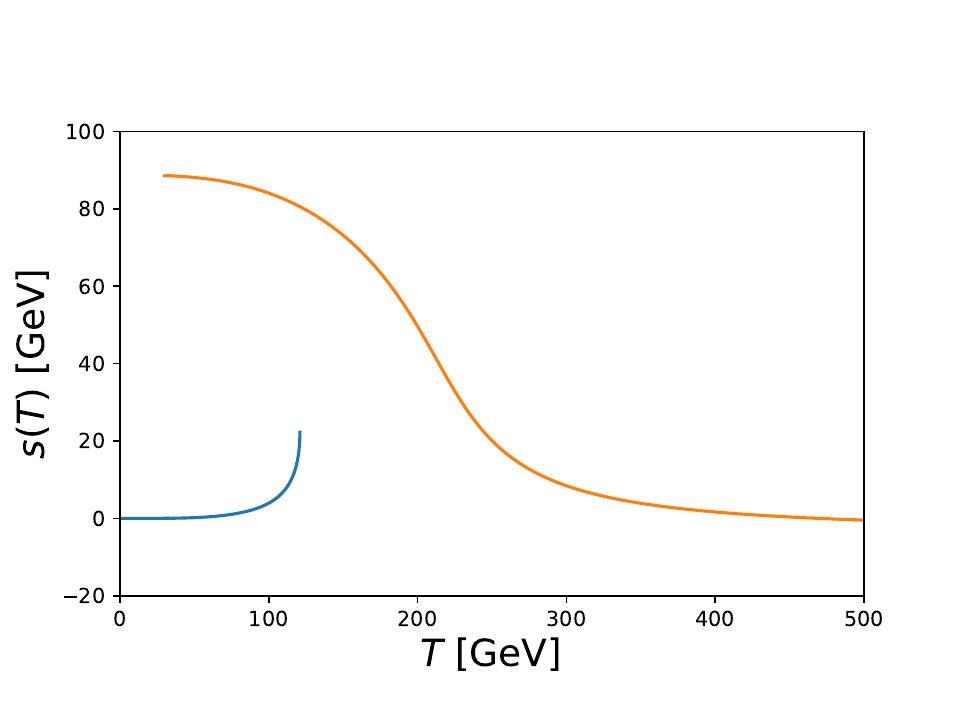}
    \caption{Evolution of VEVs of the scalars with temperature for BP8.}
    \label{fig:vevs}
\end{figure}

In our analysis, we find that phase transitions take place in two steps. At very high temperatures, the Universe resides in a phase where the singlet field develops a small but nonvanishing\footnote{At very high temperatures, the singlet scalar generally retains a nonzero VEV. Since the model lacks a discrete symmetry, explicit $\mathbb{Z}_2$-breaking terms prevent the restoration of the full symmetry along the singlet direction, inducing a VEV that is suppressed but not exactly zero. Therefore, the finite-temperature effective potential admits a nonzero minimum along the direction of the singlet field in the early Universe~\cite{Kozaczuk:2019pet}.} VEV while the Higgs VEV remains zero, corresponding to $(h,s) = (0,w_1)$. As the temperature decreases, the system undergoes a smooth crossover along the singlet direction, transitioning to a vacuum configuration $(0,w_2)$ with $w_2> w_1$. Upon further cooling, a SFOEWPT occurs from $(0,s_c) \to (v_c,0)$, with $s_c>w_2$, once the Higgs field acquires a nonzero VEV at the critical temperature $T_c$. For illustration, the phase transition pattern corresponding to BP8 is displayed in Fig.\,\ref{fig:vevs}, where the thermal evolution of the Higgs and singlet VEVs clearly shows the sequential two-step transition discussed above.

For all benchmark points examined, the second step of the transition is strongly first order along the Higgs direction, with the order-parameter ratio satisfying \(\xi_h \equiv v_c/T_c \gtrsim 1\) (see Tabs.\,\ref{tab:FOPT1} and \ref{tab:FOPT2}). In contrast, only BP9 exhibit a strong first-order transition simultaneously along both scalar-field directions. As shown in Tabs.\,\ref{tab:FOPT1} and \ref{tab:FOPT2}, the phase transition strengths \(\xi_h\) and \(\xi_s\) are comparatively weaker for BP4–BP7, which correspond to a heavier mass of the singlet scalar with \(m_{h_2} =350\ \text{GeV}\). For BP1–BP3 and BP8, where the singlet scalar is lighter with \(m_{h_2} = 200\ \text{GeV}\), the transition becomes noticeably stronger. By contrast, BP9 yields a very strong transition in both field directions. This behavior is consistent with the fact that when the additional scalar lies significantly above the electroweak scale, it effectively decouples from the thermal bath relevant for the electroweak phase transition. Consequently, its contribution to the one-loop finite temperature effective potential is suppressed, reducing its influence on the transition dynamics, as discussed earlier. 

For the zero-temperature vacuum structure considered in this work, the primary parameter governing the electroweak phase transition is the Higgs portal coupling \(\lambda_{hs}\), which serves as a free input of the model. Achieving a SFOEWPT typically requires larger values of \(\lambda_{hs}\) as the singlet-like scalar mass \(m_{h_2}\) increases. The singlet quartic coupling \(\lambda_s\) also plays an important role, as it significantly influences the shape of the scalar potential and thereby the transition dynamics.

From the perspective of direct-detection constraints, the Higgs–singlet mixing angle must be very small, which in turn implies that the associated trilinear coupling \(\mu_{hs}\) is highly suppressed. As a result, \(\mu_{hs}\) has a subdominant role on the phase-transition dynamics. The impact of the singlet trilinear coupling $\mu_3$ on the EWPT in this setup can be summarized as follows. A sizable negative value of $\mu_3$ tends to enhance the barrier in the effective potential, thereby strengthening the phase transition. Consequently, for a fixed singlet mass, a relatively smaller Higgs portal coupling is sufficient to realize a SFOEWPT. In contrast, when $\mu_3$ is positive and large, its contribution weakens the transition, requiring a larger value of $\lambda_{hs}$ to get a SFOEWPT. We also find that a SFOEWPT can be achieved even when \(\mu_3\) is set to zero. In the limit \(\mu_3 = 0\) and \(\mu_{hs} \to 0\), the model effectively reduces to a \(\mathbb{Z}_2\)-symmetric scalar extension, for which the correlation between \(\lambda_{hs}\) and the singlet mass has been extensively discussed in Refs.~\cite{Cline:2013gha,Chaudhuri:2022sis,Vaskonen:2016yiu,Kang:2017mkl}.
%It is worth noting, however, that Ref.~\cite{Ellis:2022lft} identifies an alternative scenario in which the explicit $\mathbb{Z}_2$-breaking parameters $\mu_3$ and $\mu_{hs}$ can play a significant role in strengthening the phase transition when the singlet acquires a nonzero VEV at zero temperature. Our results therefore indicate that the relevance of these parameters is strongly dependent on the underlying model and vacuum structure, and must be evaluated on a case-by-case basis.

Finally, we comment on the role of the BSM fermion $\chi$ in EWPT. Its contributions to the effective potential come from three sources. First, through thermal resummation, $\chi$ induces a temperature-dependent correction to the singlet scalar thermal mass via its Yukawa interaction (see Eq.~\eqref{eq:thFer}). In addition, $\chi$ contributes at one loop through both the CW correction and the finite-temperature part of the potential. These fermionic contributions can affect the phase transition, with their impact controlled primarily by the mixed term $2 g_\chi m_\chi s_c$ appearing in the field-dependent fermion mass squared evaluated at the critical temperature. When this combination becomes comparable to $T_c^2$, the fermionic contribution to the effective potential is maximized, thereby exerting its strongest influence on the phase-transition dynamics.

\begin{table}[ht]
\centering
\small
\setlength{\tabcolsep}{3.5pt}
\renewcommand{\arraystretch}{1.0}
\begin{tabular}{ccccccc}
\toprule
\textbf{PT param.} & \textbf{BP1} & \textbf{BP2} & \textbf{BP3} & \textbf{BP4} & \textbf{BP5} & \textbf{BP6} \\
\midrule
$(v,s)|_{T_c}^{\rm high}$ & $(0,85)$ & $(1,81)$ & $(1,90)$ & $(3,49)$ & $(4,46)$ & $(5,47)$ \\
$(v,s)|_{T_c}^{\rm low}$  & $(199,4)$ & $(196,3)$ & $(194,7)$ & $(151,4)$ & $(141,2)$ & $(143,6)$ \\
$T_c$  & 108 & 109 & 111 & 122 & 124 & 125 \\
$(\xi_h,\xi_s)$ & $(1.84,0.84)$ & $(1.79,0.72)$ & $(1.74,0.75)$ & $(1.21,0.37)$ & $(1.10,0.35)$ & $(1.10,0.33)$ \\
$(v,s)|_{T_n}^{\rm high}$ & $(0,87)$ & $(0,83)$ & $(0,92)$ & $(3,49)$ & $(4,47)$ & $(5,47)$ \\
$(v,s)|_{T_n}^{\rm low}$  & $(225,1)$ & $(221,1)$ & $(218,3)$ & $(167,2)$ & $(158,1)$ & $(156,4)$ \\
$T_n$ & 87 & 91 & 95 & 118 & 120 & 122 \\
$\alpha_n$ & 0.07 & 0.06 & 0.05 & 0.02 & 0.02 & 0.02 \\
$\beta/H_n$ & 428 & 789 & 772 & 4525 & 5902 & 8874 \\
\bottomrule
\end{tabular}
\caption{Phase transition parameters. VEVs and temperatures units of GeV.}
\label{tab:FOPT1}
\end{table}

\begin{table}[ht]
\centering
\small
\setlength{\tabcolsep}{4pt}
\renewcommand{\arraystretch}{1.05}
\begin{tabular}{cccc}
\toprule
\textbf{PT param.} & \textbf{BP7} & \textbf{BP8} & \textbf{BP9} \\
\midrule
$(v,s|_{T_c}^{\rm high}$ & $(2,50)$ & $(1,83)$ & $(0,148)$ \\
$(v,s)|_{T_c}^{\rm low}$  & $(175,6)$ & $(201,6)$ & $(232,5)$ \\
$T_c$ & 116 & 108 & 78 \\
$(\xi_h,\xi_s)$ & $(1.49,0.38)$ & $(1.85,0.71)$ & $(2.97,1.83)$ \\
$(v,s)|_{T_n}^{\rm high}$ & $(2,52)$ & $(1,86)$ & $(1,149)$ \\
$(v,s)|_{T_n}^{\rm low}$  & $(196,4)$ & $(226,2)$ & $(245,1)$ \\
$T_n$ & 107 & 86 & 43 \\
$\alpha_n$ & 0.03 & 0.06 & 0.35 \\
$\beta/H_n$ & 2030 & 719 & 587 \\
\bottomrule
\end{tabular}
\caption{Phase transition parameters. All the VEVs and temperatures are in units of GeV.}
\label{tab:FOPT2}
\end{table}

\subsection{Stochastic gravitational waves}
\begin{comment}
\end{comment}

As the temperature of the Universe drops below the critical temperature $T_c$, the electroweak-symmetric phase becomes metastable, and the transition
to the true vacuum proceeds via thermal tunneling through the potential barrier. This process occurs through the nucleation of critical bubbles of
the broken phase, whose probability is governed by the finite-temperature Euclidean (bounce) action~\cite{Linde:1981zj,Affleck:1980ac,Grojean:2006bp}.
The bounce action exponentially suppresses the tunneling rate and determines the temperature at which the phase transition effectively becomes complete.

In the finite-temperature regime relevant for the EWPT, the dominant tunneling configuration is $O(3)$ symmetric, and the corresponding bounce solution is obtained by minimizing the three-dimensional Euclidean action constructed from the finite-temperature effective potential~\cite{Linde:1977mm,Linde:1981zj}. The associated bounce equation must satisfy regularity at the bubble center and approach the false vacuum at large distances, ensuring a physically consistent nucleation process.

In this work, the bounce action and the nucleation temperature are computed numerically using the \texttt{CosmoTransitions} package~\cite{Wainwright:2011kj}, which provides a reliable and efficient framework for analyzing multi-field first-order phase transitions. This allows us to accurately determine the thermal tunneling dynamics and to characterize the EWPT relevant for our benchmark scenarios.

With the phase transition dynamics determined, the resulting evolution of the scalar fields leads to the formation and expansion of bubbles of the
broken phase, whose subsequent collisions source a stochastic gravitational-wave background. Two parameters play a central role in characterizing the GW signal from a first-order phase transition: the transition strength parameter $\alpha_n$ and the inverse duration parameter $\beta$, both evaluated at the nucleation temperature $T_n$.

The parameter $\alpha_n$ is defined as the ratio of the latent heat released during the transition to the radiation energy density,
\begin{equation}\label{eq:alphaP}
\alpha_n = \frac{\Delta \rho}{\rho_{\rm rad}},
\end{equation}
where $\rho_{\rm rad} = \pi^2 g_* T_n^4/30$ is the radiation energy density, with \( g_* \) denoting the effective number of relativistic degrees of freedom. For the SM, \( g_* = 106.75 \)~\cite{Husdal:2016haj}. The latent heat $\Delta \rho$ is computed as~\cite{Kamionkowski:1993fg,Kehayias:2009tn}
\begin{equation}\label{eq:delrho}
\Delta \rho =
\left[ V_{\rm eff}(\phi_0,T) - T \frac{d V_{\rm eff}(\phi_0,T)}{dT} \right]_{T=T_n}
-
\left[ V_{\rm eff}(\phi_n,T) - T \frac{d V_{\rm eff}(\phi_n,T)}{dT} \right]_{T=T_n},
\end{equation}
where $\phi_0$ and $\phi_n$ correspond to the field values in the false and true vacua at $T_n$, respectively. Their numerical values are summarized in Tabs.~\ref{tab:FOPT1} and~\ref{tab:FOPT2}.

The inverse duration of the phase transition is quantified by the parameter $\beta$, defined as
\begin{equation}\label{eq:betaP}
\frac{\beta}{H_n}
=
T_n \frac{d}{dT}
\left( \frac{S_E(T)}{T} \right)
\Bigg|_{T=T_n},
\end{equation}
where $H_n$ denotes the Hubble expansion rate at the nucleation temperature. The nucleation temperature $T_n$ is determined by solving the condition $S_E(T_n)/T_n \simeq 140$, which corresponds to the onset of efficient bubble nucleation~\cite{Wainwright:2011kj}.

The stochastic GW background generated by a first-order phase transition receives contributions from three main sources: bubble wall collisions, sound waves in the plasma, and magneto-hydrodynamic (MHD) turbulence. The total present-day GW energy density spectrum can be expressed as~\cite{Caprini:2015zlo}
\begin{equation}\label{eq:GWTotal}
\Omega_{\rm GW} h^2
\simeq
\Omega_{\rm col} h^2
+
\Omega_{\rm sw} h^2
+
\Omega_{\rm tur} h^2,
\end{equation}
where $h = H_0/(100\,{\rm km\,s^{-1}\,Mpc^{-1}})$ is the dimensionless Hubble parameter, with $H_0$ being the present-day Hubble constant~\cite{DES:2017txv}. The resulting GW spectrum thus provides a direct observational probe of the underlying particle physics responsible for the phase transition.

The GW contribution sourced by bubble wall collisions can be computed using the envelope approximation~\cite{Jinno:2016vai}. The resulting energy density spectrum as a function of frequency $f$ is given by
\begin{equation}\label{eq:GWcoldetails}
\Omega_{\rm col} h^2 = 1.67 \times 10^{-5} {\left(\frac{\beta }{H_n} \right)^{-2}} \left( \frac{\kappa_c \alpha_n}{1 + \alpha_n} \right)^2 \left(  \frac{100}{g^{\ast}} \right)^{1/3} \left( \frac{0.11 v^3_w}{0.42 + v^2_w} \right) \frac{3.8 \left( f/f_{\rm col} \right)^{2.8}}{1 + 2.8 \left( f/f_{\rm col} \right)^{3.8}},
\end{equation}
where $v_w$ is the bubble wall velocity and $\kappa_c$ is the efficiency factor describing the fraction of vacuum energy converted into the kinetic energy of the bubble walls. The efficiency factor is parametrized as
\begin{equation}\label{eq:kcfac}
\kappa_c = \frac{0.715 \alpha_n + \frac{4}{27} \sqrt{\frac{3 \alpha_n}{2}}}{1 + 0.715 \alpha_n}.
\end{equation}
The redshifted peak frequency associated with the collision contribution is
\begin{equation}\label{eq:PF1}
f_{\rm col} = 16.5 \times 10^{-6} \left( \frac{f_{\ast}}{\beta} \right) \left( \frac{\beta}{H_n} \right) \left( \frac{T_n}{100 \, {\rm GeV}} \right) \left( \frac{g^{\ast}}{100} \right)^{1/6} \, {\rm Hz},
\end{equation}
where the fitting function $f_\ast/\beta$ is given by~\cite{Jinno:2016vai}
\begin{equation}\label{eq:fastbybetadetails}
\frac{f_{\ast}}{\beta} = \frac{0.62}{1.8 - 0.1 v_w + v^2_w}.
\end{equation}
In the following analysis, we assume $v_w = 1$~\cite{Kamionkowski:1993fg,Espinosa:2010hh}, corresponding to relativistic bubble expansion.

A dominant contribution to the GW signal typically arises from sound waves generated in the plasma after bubble percolation~\cite{Hindmarsh:2013xza,Hindmarsh:2016lnk,Hindmarsh:2017gnf}. The corresponding GW energy density spectrum is
\begin{equation}\label{eq:GWswpart}
\Omega_{\rm sw} h^2 = 2.65 \times 10^{-6}\; \Upsilon(\tau_{\rm sw}) \left(  \frac{\beta}{H_n} \right)^{-1} v_w \left( \frac{\kappa_{\rm sw} \alpha_n}{1 + \alpha_n} \right)^2 \left( \frac{100}{g^{\ast}} \right)^{1/3} \left( \frac{f}{f_{\rm sw}} \right)^3 \left[ \frac{7}{4 + 3 \left( f/f_{\rm sw} \right)^2} \right]^{7/2},
\end{equation}
where $\kappa_{\rm sw}$ denotes the efficiency factor associated with the conversion of latent heat into bulk fluid motion~\cite{Kamionkowski:1993fg,Borah:2025hpo},
\begin{equation}\label{eq:kappasw}
\kappa_{\rm sw} =\frac{\sqrt{\alpha_n}}{0.135+\sqrt{0.98+\alpha_n}}.
\end{equation}
The peak frequency of the sound wave contribution is given by
\begin{equation}\label{eq:PF2}
f_{\rm sw} = 1.9 \times 10^{-5} \left( \frac{1}{v_w} \right) \left( \frac{\beta}{H_n} \right) \left( \frac{T_n}{100 \, {\rm GeV}} \right) \left( \frac{g^{\ast}}{100} \right)^{1/6} \, {\rm Hz}.
\end{equation}
The factor $\Upsilon(\tau_{\rm sw})$ accounts for the finite lifetime of the sound wave period and is defined as
\begin{equation}\label{eq:swtimepart}
\Upsilon(\tau_{\rm sw}) = 1 - \frac{1}{\sqrt{1+2 \tau_{\rm sw} H_{\ast}}},
\end{equation}
where $\tau_{\rm sw}$ denotes the sound wave lifetime. Following Ref.~\cite{Hindmarsh:2017gnf}, we approximate $\tau_{\rm sw} \approx R_n/\overline{U}_f$, with the mean bubble separation $R_n = (8\pi)^{1/3} v_w \beta_n^{-1}$ and the root-mean-squared fluid velocity $\overline{U}_f = \sqrt{3\kappa_{\rm sw}\alpha_n/4}$.

Finally, MHD turbulence generated in the fully ionized plasma provides an additional GW source~\cite{Caprini:2009yp}. The corresponding contribution to the GW energy density spectrum is
\begin{equation}\label{eq:GWturpart}
\Omega_{\rm tur} h^2 = 3.35 \times 10^{-4} \left( \frac{\beta}{H_n} \right)^{-1} v_w \left( \frac{\kappa_{\rm tur} \alpha_n}{1 + \alpha_n} \right)^{3/2} \left( \frac{100}{g^{\ast}}\right)^{1/3} \left[ \frac{\left( f/f_{\rm tur} \right)^3}{\left[ 1 + \left( f/f_{\rm tur} \right) \right]^{11/3} \left( 1 + \frac{8 \pi f}{h_{\ast}} \right)} \right],
\end{equation}
where $h_{\ast} = 16.5\times 10^{-6} \left( \frac{T_n}{100 \, {\rm GeV}} \right) \left( \frac{g^{\ast}}{100} \right)^{1/6} \, {\rm Hz}$ is the redshifted inverse Hubble time at GW production. The peak frequency of the turbulence contribution is
\begin{equation}\label{eq:PF3}
f_{\rm tur} = 2.7 \times 10^{-5} \frac{1}{v_w} \left( \frac{\beta}{H_n} \right) \left( \frac{T_n}{100 \, {\rm GeV}} \right) \left( \frac{g^{\ast}}{100} \right)^{1/6} \, {\rm Hz}.
\end{equation}
The turbulence efficiency factor is taken to be $\kappa_{\rm tur} = \epsilon \kappa_{\rm sw}$, where $\epsilon$ parametrizes the fraction of bulk kinetic energy converted into turbulent motion. Following previous studies~\cite{Zhou:2022mlz}, we adopt $\kappa_{\rm tur} \simeq 0.1\,\kappa_{\rm sw}$ in our numerical analysis.

\begin{figure}
    \centering
    \includegraphics[width=0.48\linewidth]{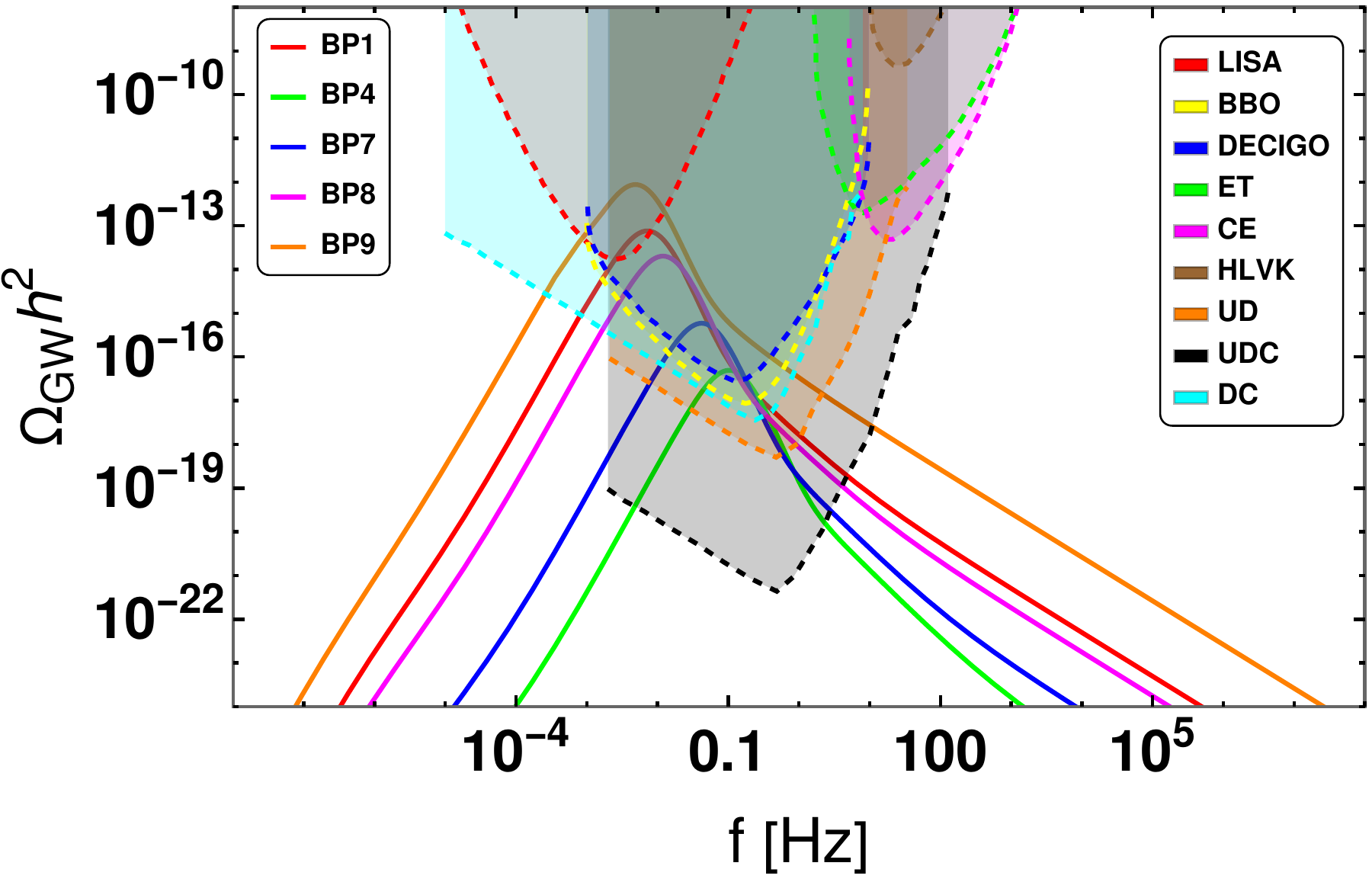}
    \includegraphics[width=0.48\linewidth]{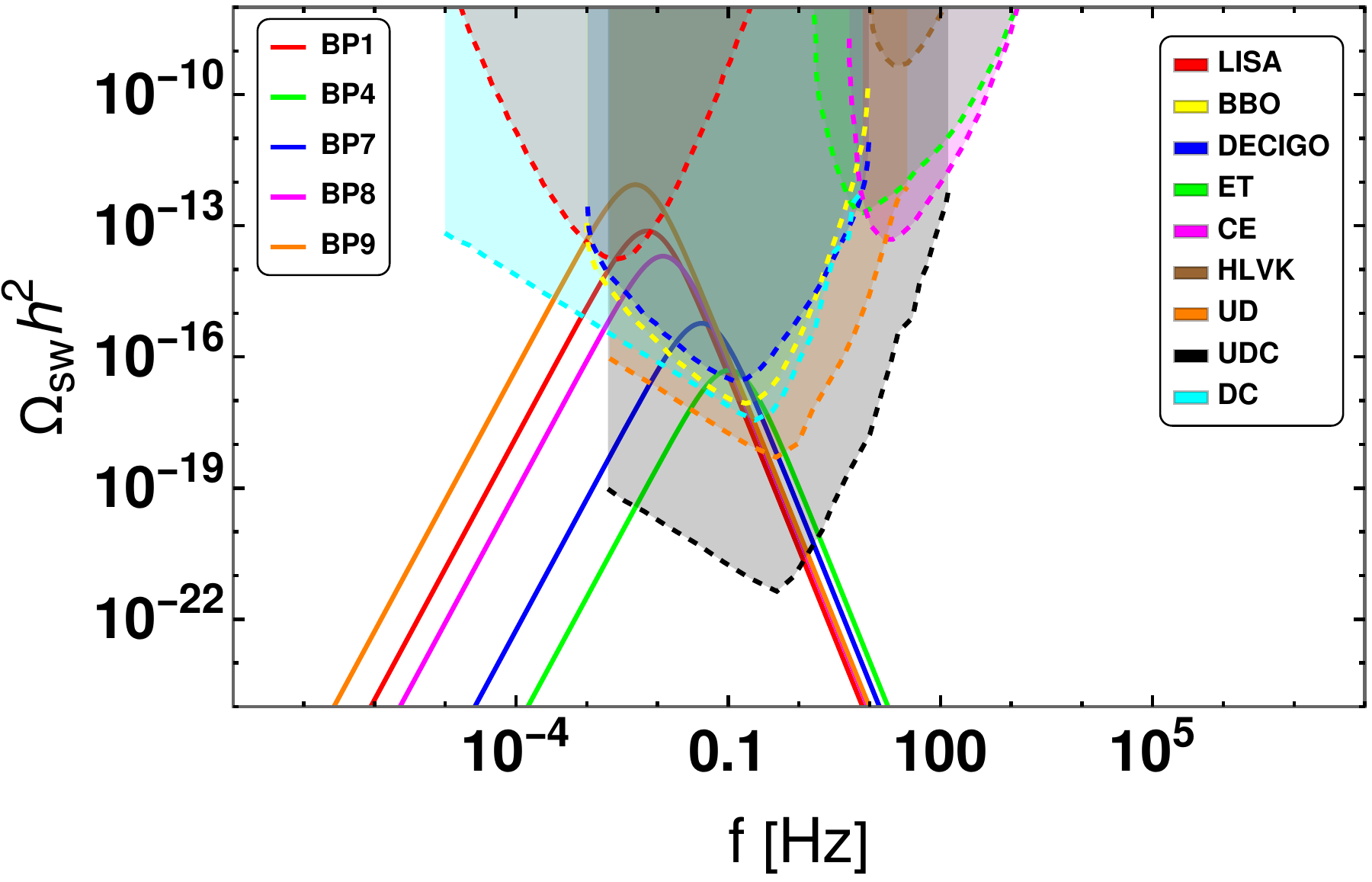}
    \includegraphics[width=0.48\linewidth]{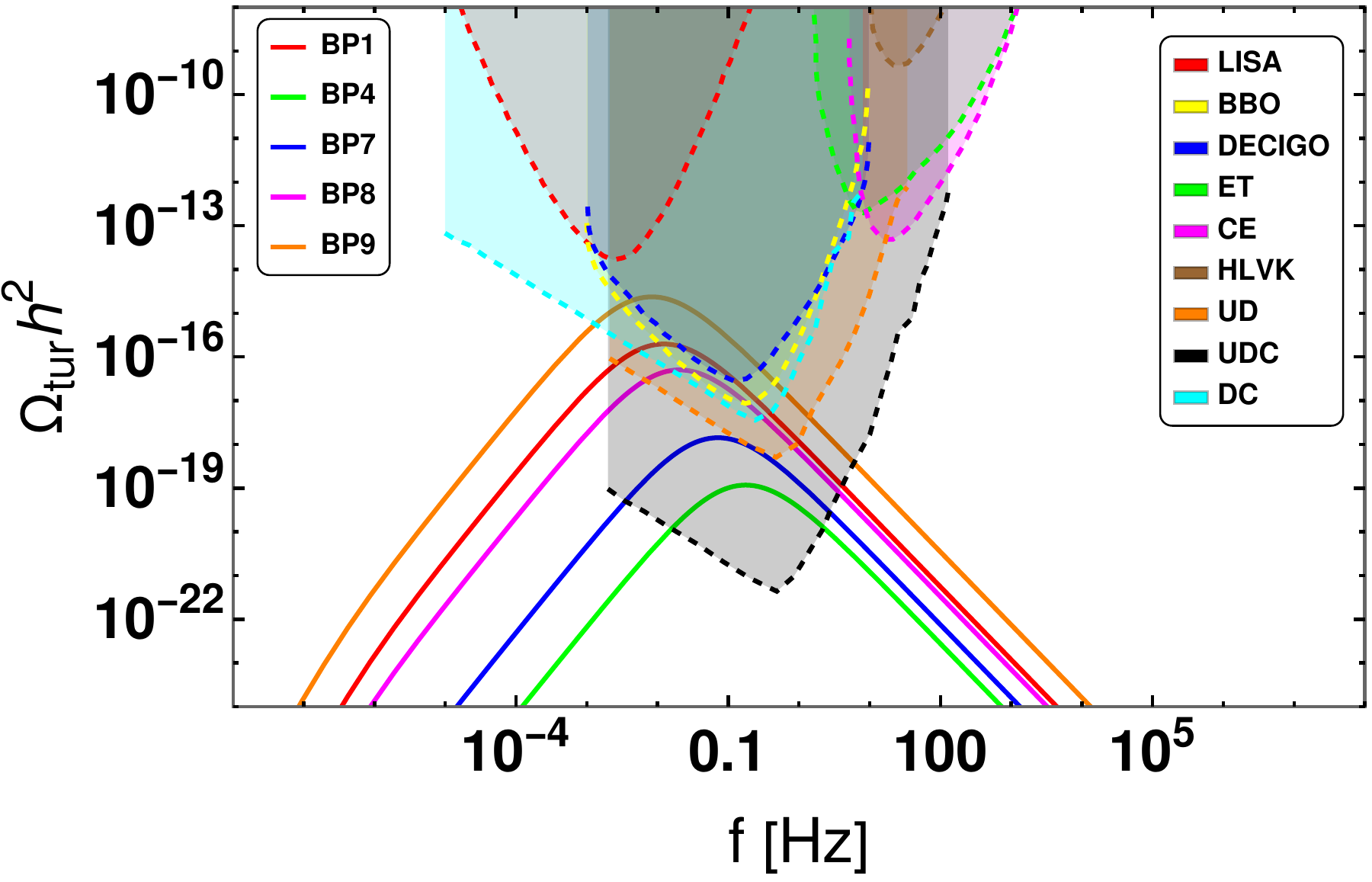}
    \includegraphics[width=0.48\linewidth]{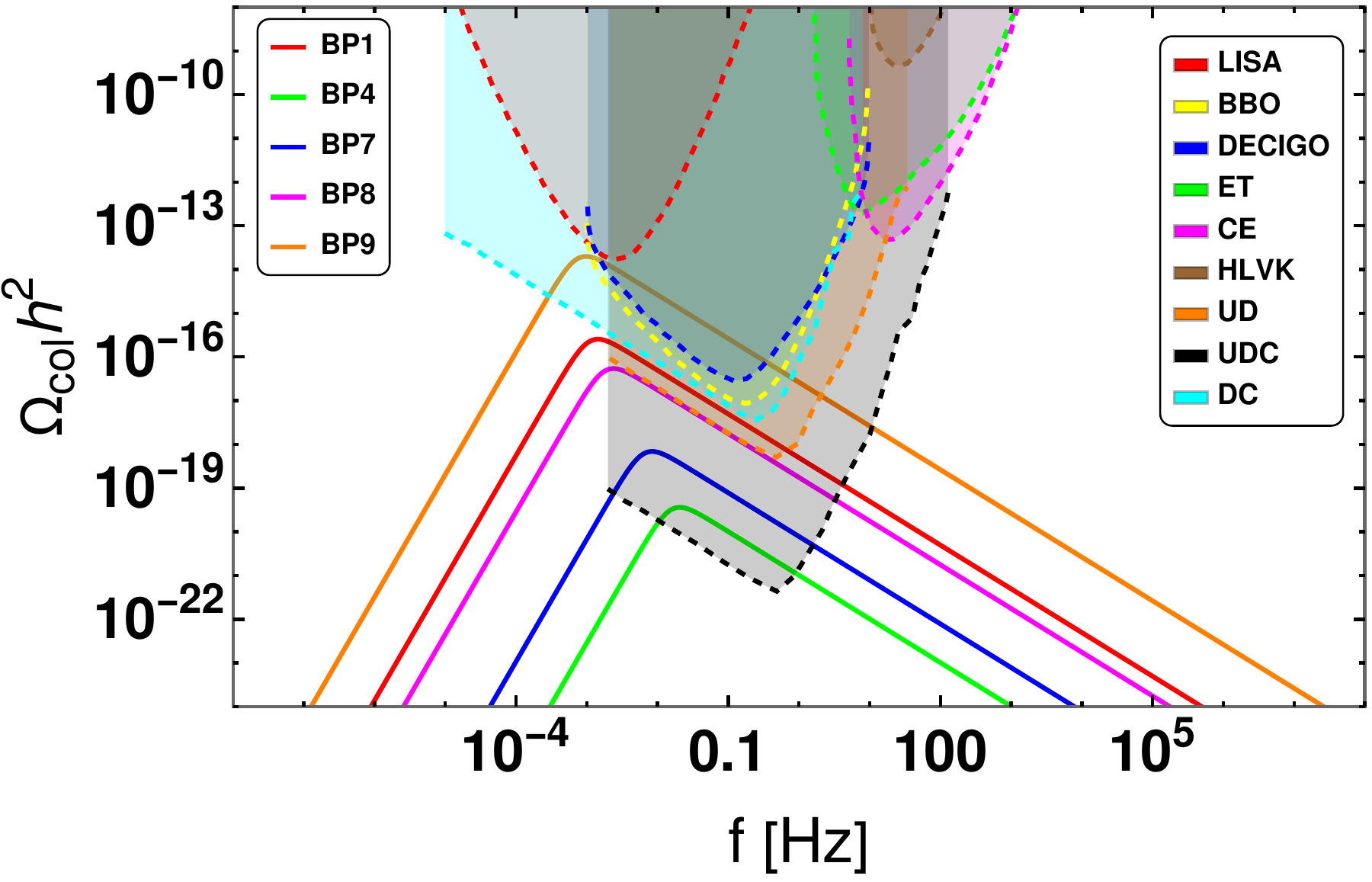}
    \caption{Variation of gravitational amplitudes with frequency. The top left panel shows the resultant GW spectrum, and the top right only the sound wave. The bottom left only shows turbulence, and the bottom right panel shows only the bubble collision. The shaded regions represent the projected sensitivity curves of future gravitational wave interferometers—LISA~\cite{LISA:2017pwj}, BBO~\cite{Yagi:2011wg}, DECIGO~\cite{Nakayama:2009ce}, Einstein Telescope (ET)~\cite{Punturo:2010zz}, Cosmic Explorer (CE)~\cite{Reitze:2019iox}, HLVK (Hanford (LIGO)-Livingston (LIGO)-Virgo-KAGRA)~\cite{LIGOScientific:2014pky,VIRGO:2014yos,KAGRA:2018plz}, ultimate-DECIGO (UD), DECIGO-corr (DC) and ultimate-DECIGO-corr (UDC)~\cite{Nakayama:2009ce} — as indicated by the legend in the inset.}
    \label{fig:gwplot}
\end{figure}

To illustrate the GW phenomenology of the model, we evaluate the spectra for the benchmark points listed in Tabs.\,\ref{tab:FOPT1} and \ref{tab:FOPT2}. The resulting predictions are shown in Fig.\,\ref{fig:gwplot}. For this analysis, we consider the “runaway’’ scenario, in which the bubbles of the broken phase accelerate rapidly and reach velocities close to the speed of light. This assumption is well motivated in the parameter space of our model, where the released vacuum energy is sufficiently large to overcome plasma friction. In this regime, GWs are generated predominantly from the sound wave as the bubble wall pushes its way through the plasma at relativistic speed \footnote{In non-runaway regimes, friction between bubble walls and the surrounding plasma prevents relativistic runaway motion, causing long-lived acoustic waves to dominate the GW production~\cite{RoperPol:2023dzg}.}. The next largest contribution comes from collisions of the bubbles expanding at high velocity. The amplitude of the GW energy density generated from the turbulence of the medium becomes subdominant. The individual contributions to the total spectrum: the sound-wave component $\Omega_{\rm sw} h^{2}$ (top right), the turbulence contribution $\Omega_{\rm tur} h^{2}$ (bottom left), and the bubble-collision envelope component $\Omega_{\rm col} h^{2}$ (bottom right).

A few general features emerge from this analysis. First, since the peak frequency scales inversely with the critical temperature, benchmarks with larger $T_c$ naturally produce spectra shifted toward higher frequencies, whereas those with lower $T_c$ give rise to peaks lying in the optimal sensitivity window of space-based interferometers. Second, the overall amplitude of the spectrum increases with the strength of the phase transition, parametrized by $\alpha$, and conversely decreases with increasing $\beta/H_n$. The strength of the phase transition and hence, the amplitude of the resultant gravitational wave increases as $T_c - T_n$ increases, which signifies how supercooled the universe was during the bubble nucleation. The GW spectrum tends to become less pronounced as the scalar mass increases, as can be observed from Fig.\,\ref{fig:gwplot} where BP9 with $m_{h_2} = 70$ GeV produces a much higher peak than BP4 with $m_{h_2} = 350$ GeV (see Tab.\,\ref{tab:BPs}).

 Several of our benchmark points lead to GW spectra with amplitudes that fall within the projected sensitivity of next-generation experiments. In particular, the spectra corresponding to BP9 exhibit the largest signal, featuring a peak around $f \sim 0.01~\text{Hz}$ that lies well within the expected discovery reach of LISA~\cite{LISA:2017pwj}, BBO~\cite{Yagi:2011wg}, and DECIGO~\cite{Nakayama:2009ce}. Moreover, this framework can also be probed by proposed experiments targeting the higher-frequency band $f \sim 1\text{--}10~\text{Hz}$, such as UD and UDC, providing an additional avenue for testing the model. 
 In fact, the GW amplitude is highest for BP9, since both $T_c - T_n$ and the corresponding strength parameter $\alpha_n$ take their largest values in this benchmark. By contrast, BP4 produces a significantly weaker signal that remains below the projected reach of most planned detectors. Nevertheless, it can still be probed by Ultimate-DECIGO and correlated-DECIGO~\cite{Nakayama:2009ce}. On the other hand, the sensitivities of HLVK~\cite{LIGOScientific:2014pky,VIRGO:2014yos,KAGRA:2018plz}, ET~\cite{Punturo:2010zz}, and CE~\cite{Reitze:2019iox} (see the caption of Fig.\,\ref{fig:gwplot} for the nomenclature) are insufficient to test this scenario.

The wide range of frequencies present in the predicted GW spectra for this model, in association with the varied sensitivity curves of existing and future GW detectors, underscores a key point: different GW observatories are complementary in their ability to probe such an extended scalar sector model that predicts a SFOEWPT.

\section{Summary and conclusions}\label{sec:conclusion}

In this work, we revisited a minimal and well-motivated extension of the Standard Model in which a SM gauge-singlet Dirac fermion $\chi$ constitutes the dark matter candidate and interacts with the visible sector through a real SM gauge-singlet scalar $s$. The scalar couples to the SM Higgs doublet via cubic and quartic Higgs-portal interactions and mixes with the Higgs after electroweak symmetry breaking. A central structural feature of this setup is that the singlet scalar does not acquire a vacuum expectation value at zero temperature. Consequently, the Higgs-singlet mixing is controlled entirely by the dimensionful portal parameter $\mu_{hs}$, while the quartic portal coupling $\lambda_{hs}$ governs the finite-temperature scalar potential.

This separation of roles has important phenomenological consequences. In particular, the coupling $\lambda_{hs}$, which drives the strength of the EWPT, is largely decoupled from the parameters controlling collider observables and dark-matter direct-detection rates. This allows the model to avoid the tension typically encountered in singlet-extended scenarios between achieving a strong first-order EWPT and satisfying experimental constraints.

We first imposed perturbative unitarity and vacuum-stability requirements on the scalar potential. Collider and electroweak precision data further restrict the viable parameter space. Using the combined ATLAS Higgs signal-strength measurement, we obtained an approximate $95\%$ confidence-level bound $|\sin\theta|\lesssim 0.24$, largely independent of the heavy-scalar mass. Direct searches for heavy scalar resonances decaying into $ZZ$, $WW$, and $h_1h_1$ final states further constrain the mixing angle to $|\sin\theta|\lesssim\mathcal{O}(0.1)$ for $m_{h_2}\lesssim 600$~GeV. Taken together, these results confine the model to a region with small but non-negligible scalar mixing.

We then carried out a detailed analysis of the dark-matter phenomenology. Thermal freeze-out of $\chi$ is dominated by scalar-mediated annihilation channels such as $\chi\bar{\chi}\to h_i h_j$ $(i,j=1,2)$, with additional contributions from annihilation into SM final states near scalar resonances. The observed relic abundance selects viable regions of parameter space with distinct characteristics. In the Higgs-resonance region, achieving the correct relic density typically requires relatively large mixing angles. In contrast, near the singlet-resonance regime $m_\chi\simeq m_{h_2}/2$, the observed abundance can be obtained over a wider range of $\sin\theta$ and $g_\chi$ values. For $m_\chi\gtrsim m_{h_2}$, the relic density is reproduced only for comparatively large Yukawa couplings, with a reduced dependence on the mixing angle.

Direct-detection constraints play a dominant role in shaping the allowed parameter space. The spin-independent dark-matter-nucleon scattering cross section depends sensitively on the scalar mixing angle and the heavy-scalar mass. Current bounds exclude large mixing angles, while values as small as $\sin\theta\sim10^{-3}$ remain well below present experimental sensitivity even for ${\cal O}(1)$ Yukawa couplings. In addition, destructive interference between the two scalar propagators gives rise to blind-spot regions where the scattering cross section is strongly suppressed. A combined analysis of relic density and direct-detection constraints shows that several benchmark points lie close to the $2\sigma$ exclusion boundary of current experiments.

From the collider perspective, the production rates for non-standard di-scalar final states are generally suppressed due to the small Higgs-singlet mixing. For benchmark points BP1-BP7, both $pp\to h_1h_2$ and $pp\to h_2h_2$ cross sections lie well below the femtobarn level. An exception is BP8, for which the rates reach the sub-femtobarn level, making it the most promising benchmark for future collider sensitivity. In BP9, the enhancement of di-scalar production is primarily driven by phase-space effects associated with the lighter $h_2$ mass rather than by an increased mixing angle.

An important observation is that the enhancement observed for BP8 is correlated with a resonant configuration in which the dark-matter mass lies close to $m_{h_2}/2$. By explicitly varying $m_\chi$ near this resonance, we find that the scalar mixing angle can increase to moderately larger values while remaining consistent with current collider and direct-detection constraints. As a result, the associated production rate of the heavier scalar is enhanced and can approach the femtobarn level. Such configurations may therefore become accessible at the HL-LHC and constitute well-motivated targets for future high-energy hadron colliders such as the FCC-hh. A detailed collider-level analysis is left for future work.

The central novelty of this study lies in the interplay between the dark-matter sector and the electroweak phase transition. Using \texttt{CosmoTransitions}, we demonstrated that the model exhibits a robust two-step phase-transition pattern. For all benchmark points considered, the second step is strongly first order along the Higgs direction, satisfying $\xi_h = v_c/T_c \gtrsim 1$. The dominant contribution to strengthening the phase transition arises from the scalar sector, while the fermionic dark matter affects the finite-temperature potential only through Yukawa-induced loop corrections. Importantly, the small scalar mixing required by direct-detection constraints does not weaken the phase transition.

We also investigated the stochastic gravitational-wave signals associated with the strong first-order electroweak phase transition. Several benchmark points yield spectra with peak amplitudes within the projected sensitivities of future space-based interferometers such as LISA, DECIGO, u-DECIGO, and BBO. In particular, BP9 produces the strongest signal, with a peak frequency around $f\sim10^{-2}$~Hz, well aligned with the optimal sensitivity range of these experiments.

To summarise, the main outcomes of this study are:
\begin{itemize}
  \item In conventional singlet-scalar extensions where the singlet field acquires a vacuum expectation value, realizing a strong first-order EWPT typically requires a large Higgs--portal coupling. In the presence of a singlet VEV, this leads to sizable Higgs-singlet mixing, which is strongly constrained by direct-detection experiments and collider data. The key novelty of this work lies in disentangling $\lambda_{hs}$ and $\sin\theta$ by forbidding a singlet VEV at zero temperature and introducing the dimensionful parameter $\mu_{hs}$ to control the mixing independently.

  \item The absence of a singlet VEV allows the Higgs-portal quartic coupling $\lambda_{hs}$ to be sizable while maintaining a small scalar mixing angle. This structure generates a tree-level barrier in the finite-temperature potential, enabling a strong first-order EWPT, while simultaneously suppressing direct-detection rates and deviations in Higgs observables. For suitable parameter choices, the model satisfies all current experimental constraints and predicts GW signals testable by next-generation detectors.
\end{itemize}

A more comprehensive collider study, including refined projections for future facilities, would further clarify the connection between the GW–favoured region and scalar-resonance or di-Higgs signatures. Extensions of the framework with additional degrees of freedom could further enrich the phase-transition dynamics while preserving the close connection between dark matter, collider phenomenology, and gravitational-wave signatures established here.

\section*{Acknowledgment}
The authors thank Prof. Debajyoti Choudhury for useful discussions and Dr. Yikun Wang for his helpful communication. JD acknowledges SERB, Government of India, for the National Postdoctoral Fellowship (NPDF) grant with File No. PDF/2023/001540. SN
acknowledges funding from Anusandhan National Research Foundation (formerly Science and
Engineering Research Board (SERB)), Govt. of India with grant number SUR/2022/001404 of SURE scheme. TS is supported by the DST, Government of India through the DST INSPIRE Faculty Fellowship (DST/INSPIRE/04/2024/004616). The work is also supported by the National Natural Science Foundation of China under Grant Nos. 12475094, 12135006, and 12075097. 

\appendix
\section{Coefficients of $V_{\rm ct}(h,s)$}\label{app:counter-term}
The renormalization conditions for calculating the coefficients of $V_{\rm ct}(h,s)$ given in Eq.~\eqref{eq:ctpot} are
\begin{eqnarray}
 \left\{\frac{d}{dh},\frac{d}{ds}, \frac{d^2}{dh ds},\frac{d^2}{dh^2},\frac{d^2}{ds^2}\right\} \Big(V^{\rm CW}_{1-\rm loop}(h,s,T)+V_{\rm ct}(h,s)\Big) \Big|_{h=v,s=0,T=0}=0.
 \end{eqnarray}
 Using the above conditions, the coefficients can be written as
\begin{eqnarray}
   && \delta\mu_h^2= -\frac{1}{2v}\left(3 \frac{dV^{\rm CW}}{d h}  -v \frac{d^2V^{\rm CW}}{d h^2} \right)\Bigg|_{h=v,s=0},\\
   &&\delta\lambda_h = -\frac{1}{2v^3}\left(v \frac{d^2V^{\rm CW}}{d h^2} -\frac{dV^{\rm CW}}{d h} \right)\Bigg|_{h=v,s=0}, \\
   && \delta\mu_s^2= - \frac{d^2V^{\rm CW}}{d s^2}\Bigg|_{h=v,s=0} ,\\
   &&\delta\mu_{hs} =  -\dfrac{1}{v}\frac{d^2V^{\rm CW}}{dh d s}\Bigg|_{h=v,s=0},  \\
    &&\delta a_1 = \frac{1}{2} \left(v\dfrac{d^2V^{\rm CW}}{dh ds}-2\dfrac{dV^{\rm CW}}{ ds}\right)\Bigg|_{h=v,s=0}.
\end{eqnarray}
Here, we have used the shorthand notation $V^{\rm CW}\equiv V^{\rm CW}_{1-\rm loop}(h,s,T=0)$.

\label{Bibliography}
\bibliographystyle{JHEP}
\bibliography{Refs}
\end{document}